\documentclass[twocolumn,tighten,twocolappendix]{aastex631}

\def\purple#1 {{\textcolor{purple}{#1}}\ }
\def\red#1 {\textcolor{red}{#1}}
\def\blue#1 {{\textcolor{blue}{#1}}\ }
\definecolor{forestgreen}{rgb}{0.13, 0.55, 0.13}
\def\zy#1 {\textcolor{forestgreen}{ zy: #1}}

\shorttitle{AASTeX v6.3.1 Sample article}
\shortauthors{Fei Li et al.}
\graphicspath{{./}{figures/}}

\begin{document}

\title{Dense gas properties and star formation in M 82}

\author{Fei Li$^{\dagger}$}
\email{$\dagger$:lifei@nju.edu.cn}
\affiliation{School of Astronomy and Space Science, Nanjing University, Nanjing 210093, People’s Republic of China}
\affiliation{Key Laboratory of Radio Astronomy, Chinese Academy of Sciences,  10 Yuanhua Road, Nanjing, JiangSu 210033, China}

\author{Zhi-Yu Zhang$^{\dagger}$}
\email{$\dagger$:zzhang@nju.edu.cn}
\affiliation{School of Astronomy and Space Science, Nanjing University, Nanjing 210093, People’s Republic of China}
\affiliation{Key Laboratory of Radio Astronomy, Chinese Academy of Sciences,  10 Yuanhua Road, Nanjing, JiangSu 210033, China}


\author{Junzhi Wang$^{\dagger}$}
\email{$\dagger$:junzhiwang@gxu.edu.cn}
\affiliation{School of Physical Science and Technology, Guangxi University, Nanning 530004, People’s Republic of China}

\author{Gan Luo}
\affiliation{Institut de Radioastronomie Millimetrique, 300 rue de la Piscine, Domaine Universitaire de Grenoble, 38406, Saint-Martin d’Hères, France}



\begin{abstract}
High-density molecular gas plays a vital role in supporting star formation
within galaxies. However, traditional tracers of dense gas, such as the low-$J$
transitions of HCN and HCO$^+$, are predominantly optically thick. This
characteristic presents a significant challenge in accurately estimating the
column density of dense gas and mapping its spatial distribution. Optically-thin
tracers, including HC$_3$N (10-9) and isotopologues of HCN, HCO$^+$, HNC (1-0),
among others, emerge as better tracers, by enabling more precise measurements
and analyses of the dense gas in galaxies. Here, we present the first
high-resolution ($\sim$3.8$''$, corresponding to $\sim$65 pc) molecular line
observations of the nearby starburst galaxy M~82 with IRAM Northern Extended
Millimeter Array (NOEMA). Notably, HC$_3$N (10-9) and H$^{13}$CN (1-0) emission
lines are brighter at the northeast part (NE lobe) than those at the southwest
part (SW lobe) of M~82, suggesting a higher accumulation of dense gas at the NE
lobe. The spatial distributions of H41$\alpha$ and 3-mm continuum emission
indicate that starburst activity varies across the central starburst disk, with
more intense star formation occurring at the SW lobe compared to the NE lobe.
The average optical depths of HCN and HCO$^+$ (1-0) across the
triple-peaked regions exhibit significant variation, with the highest values
observed at the NE lobe. Our results suggest a potential evolutionary sequence
in M~82, where the distributions of star formation and dense gas appear to be
partially decoupled --- a phenomenon that classical dense gas tracers cannot
adequately probe.



\end{abstract}

\keywords{galaxies:active(69) --- galaxies:ISM(113) --- galaxies:Starburst(132)}


\section{Introduction} \label{sec:intro}

Dense molecular gas, which is often traced by lines of molecules with high
dipole moment, such as HCN, HCO$^+$, HNC, and CS, is directly related to star
formation in galaxies
\citep{Gao2004b,Kennicutt2012,Usero2015,Bigiel2016,Jimenez2019}. Observations of
dense molecule gas within the central region of starburst galaxies provide an
ideal template to study not only star-forming structures, but also the evolution
of the dense interstellar medium (ISM) around galactic nuclei. 
Measurements of
the spatial distribution of HCN, HCO$^+$, HNC, and CS lines towards nearby
galaxies have resulted in a better understanding of the physical properties,
kinematics, and mass distribution of the dense gas in these regions
\citep{Martin2015,Querejeta2019,Callanan2021,Eibensteiner2022,Li2024}.  

However, the commonly adopted dense gas tracers are normally optically thick,
which could jeopardise the accurate determination of gas masses and density
structure of the dense-phase ISM
\citep{Narayanan2012,Bolatto2013,Jimenez2017,Cormier2018}. In contrast to the
routinely-adopted dense gas tracers, HC$_3$N, H$^{13}$CN, H$^{13}$CO$^+$, and
HN$^{13}$C lines also have high critical densities but are usually optically
thin (mostly due to their lower abundance), which can trace densest parts of
molecular clouds
\citep{Wang2014,Wang2016,Jimenez2017,Krieger2020,Li2020,Li2022}. 

Optically thin dense gas tracers are challenging to be observed due to the lower
abundances and smaller volumes, resulting in a significantly reduced line
brightness. Only a few detections of such lines have been reported in nearby
galaxies, which are either limited within (ultra)luminous infrared galaxies
(U/LIRGs) and starburst galaxy centers or take global properties of galaxies as
a whole \citep{Wang2014,Wang2016,Tunnard2015,Li2020}. Following advancements in
sensitivity of the single-dish telescopes and interferometers makes the
detection of the spatial distribution of weak molecular lines feasible in
external galaxies, such as HC$_3$N
\citep{Lindberg2011,Jiang2017,Mills2018,Li2024} and rarer isotopologues of HCN
and HCO$^+$.




M 82 is one of the nearest starburst systems, at a distance 
of 3.5 Mpc \citep[1\arcsec\ $\approx$ 17 pc,][]{Karachentsev2006}.
It is an edge-on \citep[$i \approx$ 80$^{\circ}$; e.g.][]{McKeith1993} prototypical starburst galaxies \citep[star formation rate (SFR) $\sim$ 10 M$_{\odot}$ yr$^{-1}$,][]{deGrijs2001,Li2021}. Its central starburst, which covers a region $\sim$ 1 kpc \citep{OConnell1978}, is observed from X-ray to millimeter and radio emission. Studies of CO \citep{Walter2002,Mao2000,Taylor2001,Leroy2015} and dense gas tracers, such as HCN, HCO$^+$, HCN, \citep{Fuente2005,Naylor2010,Ginard2015,Krieger2021} indicate that the central concentration of molecular gas and the intense ultraviolet (UV) field have influenced the physical properties, kinematics and chemistry of molecular gas. Our early IRAM 30-m observations of dense gas tracers (including isotopologues) revealed significant spatial variations in the I(HCN)/I(H$^{13}$CN) and I(HCO$^+$)/I(H$^{13}$CO$^+$) ratios along the major axis of M 82 \citep{Li2022}. Notably, these ratios are lower in the central region compared to the surrounding disk. This prompts two key questions: (1) How is dense gas linked to star formation at cloud scales in M 82? (2) What is the evolutionary stage of the nuclear starburst in its central 1 kpc region?


In this work, we present new NOEMA observations towards the central starburst disk of M 82, including 
3mm continuum, H41$\alpha$, HC$_3$N (10-9), and the cloud-scale extragalactic map of isotopologues of HCN (1-0), HCO$^+$ (1-0), and HNC (1-0). Our goal is to reveal the physical properties of the molecular clouds within the nuclear starburst of M 82 and explore the evolutionary stage of the nuclear starburst along the major axis.
The paper is organized as follows. We describe the observations and the data reduction in Section \ref{sec:observation}. Section \ref{sec:results} presents line emission distributions of 3mm continuum and line emissions followed by an analysis and discussion in Section \ref{sec:discussion}. Finally, our conclusions are summarized in Section \ref{sec:summary}.





\section{Observations and data reduction}\label{sec:observation}
The NOrthern Extended Millimetre Array (NOEMA) was used to observe the molecular lines at 3mm on the central region of M 82 (project ID: W20BT; PI: Fei Li). 
The observations were carried out using the single-field observing mode in April 2021 in the D configuration, with baselines between 32 and 288 m. The spectral lines are observed using the 3mm band receiver with the PolyFiX. 
We tuned the upper sideband to cover H$^{13}$CN (1-0), H$^{13}$CO$^{+}$ (1-0), HN$^{13}$C (1-0), H$^{15}$NC (1-0), H41$\alpha$, and HC$_3$N (10-9), with a spectral resolution of 2 MHz. 
The flux calibrators included LKHA101, MWC349, and 3C84. 1044+719 was used as the phase and amplitude calibrator. Bandpass calibrators are 3C84, 2013+370, and 0240-014. The total telescope time is 8 hours. A summary of these three observations is shown in Table \ref{tab:observation}.


The calibration of the RAW data was performed using the standard reduction pipeline in GILDAS/CLIC\footnote{\url{https://www.iram.fr/IRAMFR/GILDAS/}}. The image of the calibrated visibilities was done using the \texttt{tclean} task with a Briggs weighting (robust = 0.5) in the Common Astronomy Software Applications \citep[CASA, version 5.8.0,][]{CASA2007}. 
The final clean cube achieved an effective velocity resolution of 10 km s$^{-1}$, and the typical rms noise level is 1.05 mJy/beam. Then, we use the task \texttt{UVconsub} to subtract the continuum. The synthesized beam size is 3.8\arcsec\ $\times$ 3.0\arcsec\ with a position angle of 36$^{\circ}$. 
The primary beam is about 56\arcsec at 89~GHz.  
All images shown in this paper are prior to primary beam correction, while all measured fluxes are corrected for the primary beam attenuation.

\begin{table*}
\centering

\caption{\textbf{Observational parameters}}
\label{tab:observation}
\begin{tabular}{lllll}
\hline
\hline
Project-ID           &  W20BT\\ 
Dates                & April 1, 3, 4, and 14, 2021    \\
Total observing time &8 hours   \\   
Observed central frequency& 89.805345 GHz  \\
Flux calibrator   & LKHA101; MWC349; 3C84\\        
Bandpass calibrator  &3C84; 2013+370; 0240-014\\
phase calibrator    & 1044+719 \\ 
\hline
\bf{Spectral line parameters:}\\
 Line           &fre. (GHz)     &1$\sigma$ (mJy/beam)\\
\hline
contunuum          &82.082&0.3\\
HC$^{15}$N (1-0)   &86.055& 1.3  \\
H$^{13}$CN (1-0)   &86.340&   0.8 \\
HCN (1-0)          & 88.632& 1.4  \\
H$^{13}$CO$^+$(1-0)&86.754& 0.9\\
HN$^{13}$C (1-0)   &87.091& 1.1\\
H$^{15}$NC (1-0)   &88.866&0.8\\
HCO$^+$ (1-0)      & 89.189&1.8\\
HNC (1-0)          & 90.663 &1.3\\
HC$_3$N (10-9)     &90.979 &1.1\\
H41$\alpha$        &92.034&1.1\\

\hline

\end{tabular}\\
     
\begin{minipage}{1.35\columnwidth}
 \vspace{1mm}
{\bf Notes:} The continuum rms is for the whole bandpass, excluding the strong spectral lines. The 1$\sigma$ rms values are for all lines with a resolution of 5 km s$^{-1}$. 
 \end{minipage}

\end{table*}

\begin{figure}
\includegraphics[scale=0.17]{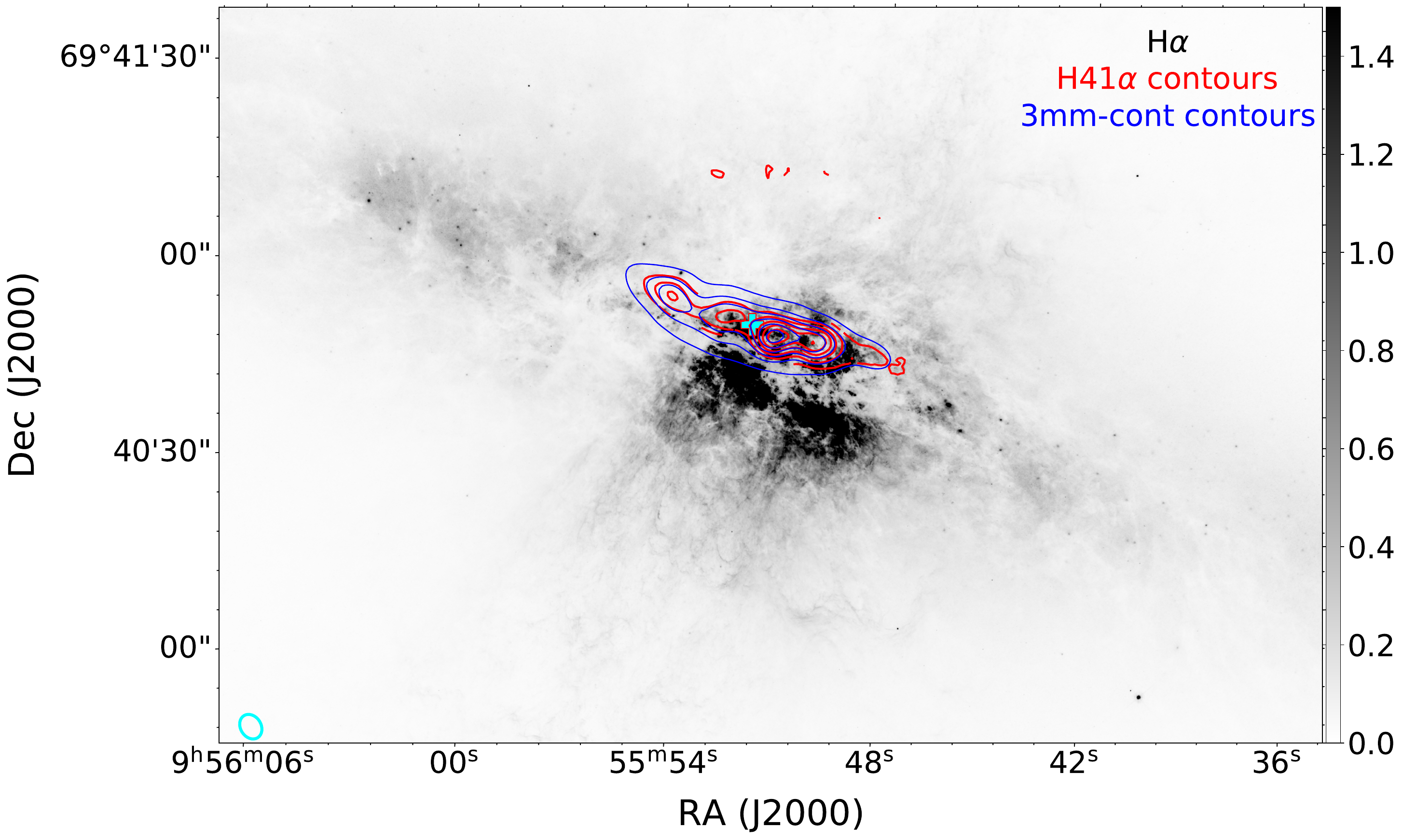}

\caption{H${\alpha}$ emission image obtained with Hubble Space Telescope (HST) \citep{Mutchler2007}. Contours levels are 0.09 Jy beam $^{-1}$ km s$^{-1}$to 1.29 Jy beam $^{-1}$ km s$^{-1}$by steps of 0.24 Jy beam $^{-1}$ km s$^{-1}$for H41$\alpha$ (red lines), 0.006 Jy/beam to 0.052 Jy/beam by steps of 0.009 Jy/beam for 3mm continuum (blue lines). The beam is shown at the bottom left corner of the panel.}
\label{fig:Halpha}
\end{figure}

\begin{figure*}
\includegraphics[scale=0.45]{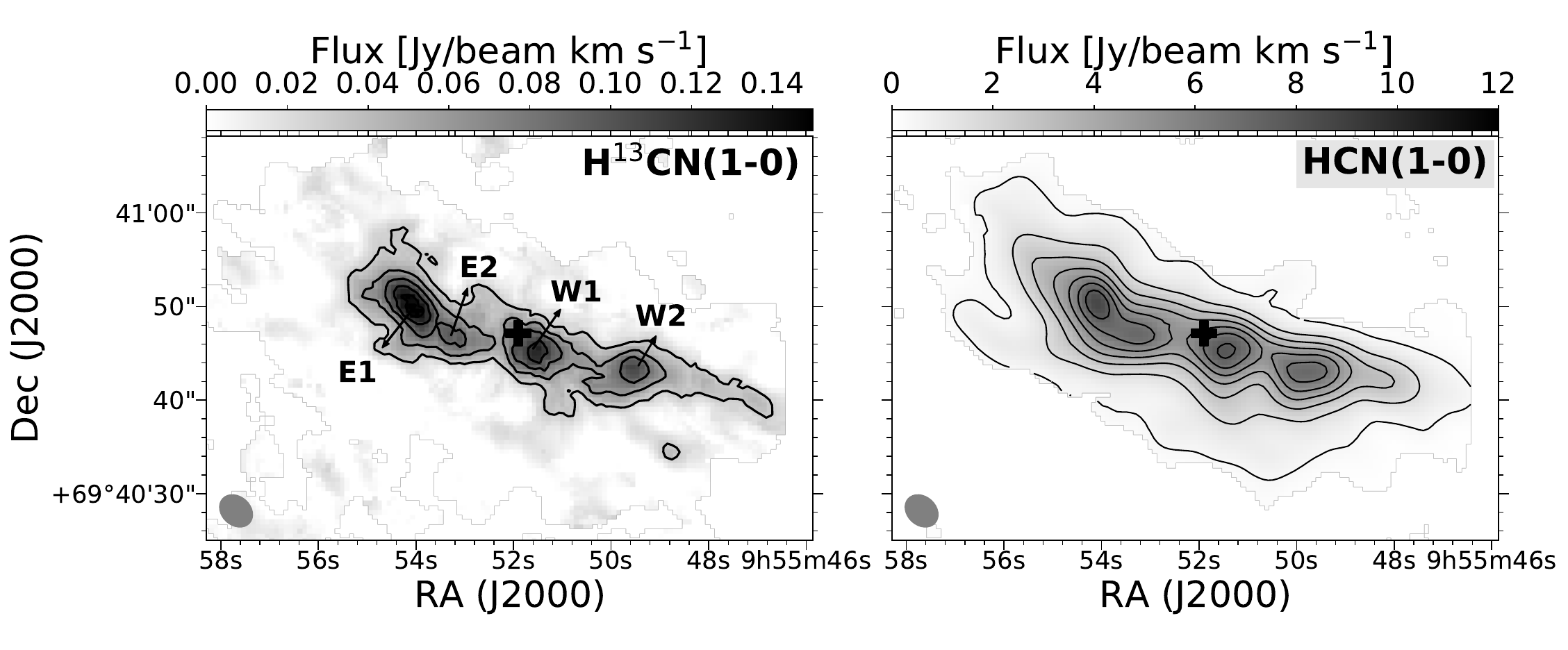}
\includegraphics[scale=0.45]{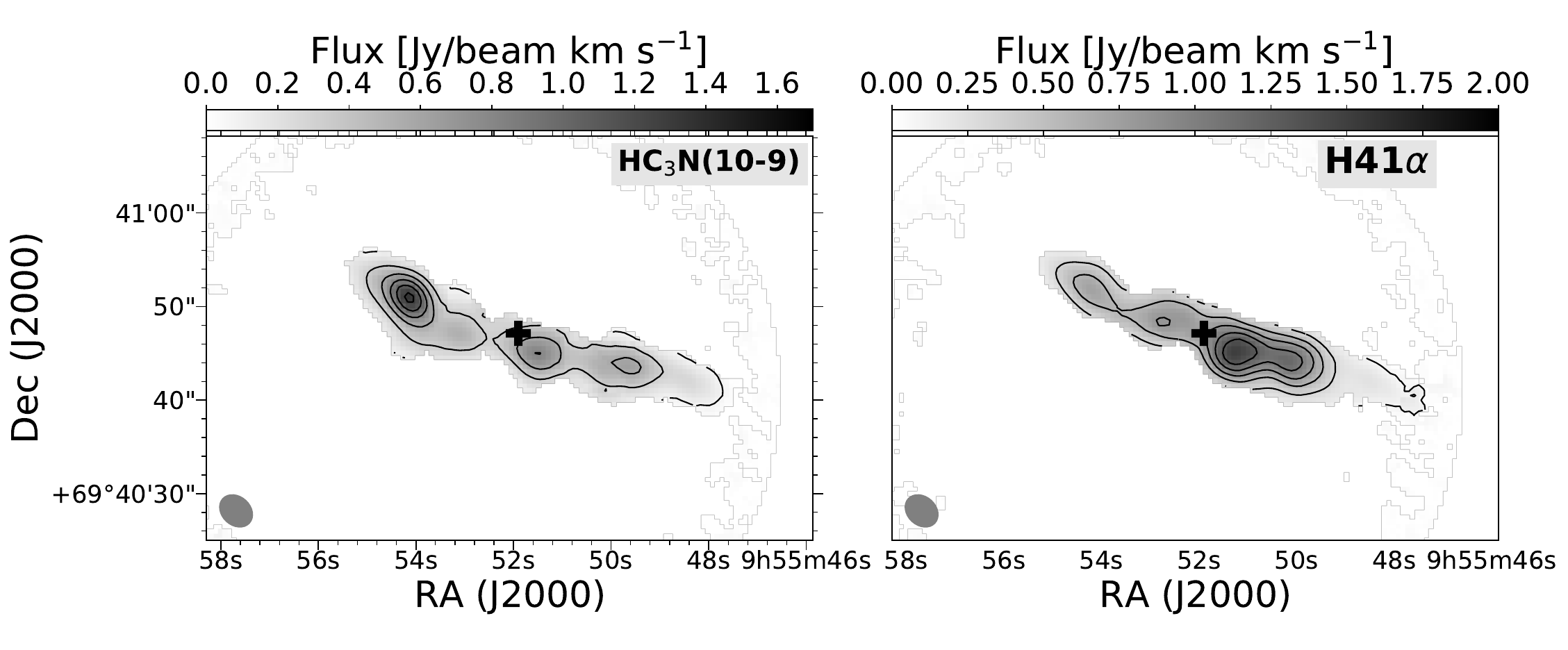}
\caption{Line-integrated intensity maps of H$^{13}$CN (1-0), HCN (1-0), HC$_3$N (10-9), and H41$\alpha$. The beam is shown at the bottom left corner of the panels as a shaded ellipse. The dynamical center of the galaxy is marked by the central filled plus. The contours levels start from 0.03 Jy beam $^{-1}$ km s$^{-1}$to 0.17 Jy beam $^{-1}$ km s$^{-1}$in steps of 0.035 Jy beam $^{-1}$ km s$^{-1}$for H$^{13}$CN, 0.42 Jy beam $^{-1}$ km s$^{-1}$to 7.62 Jy beam $^{-1}$ km s$^{-1}$in steps of 1.2 Jy beam $^{-1}$ km s$^{-1}$for HCN,  0.09 Jy beam $^{-1}$ km s$^{-1}$to 1.29 Jy beam $^{-1}$ km s$^{-1}$in steps of 0.24 Jy beam $^{-1}$ km s$^{-1}$for HC$_3$N. The contour levels of H41$\alpha$ are similar to Figure \ref{fig:Halpha}. First contour is at 3 $\times \sigma$ level.}
\label{fig:HC3N_H41a_HCN_continuum}
\end{figure*}


\section{Results}
\label{sec:results} 

To analyze the spatial distribution of the dense gas in the central starburst of M 82, we fit each pixel’s spectrum using a single component Gaussian profile to obtain the full-width-at-half maximum (FWHM) and central velocity. Then, we generate the integrated intensity map (moment 0) of each line using the \texttt{moment()} method from the Python package \texttt{SpectralCube} according to the FWHM of each pixel. To improve the signal-to-noise ratio (S/N) of the final images, we applied a masking routine with 2$\sigma$ clipping before producing the moment 0 maps. Since HCO$^+$ (1-0) is the strongest spectral feature that we observed and the central velocity and line width of HCN (1-0), HCO$^+$ (1-0) and HNC (1-0) are almost the same, we use HCO$^+$ (1-0) as a reference for all isotopic lines. The moment 0 maps of H$^{13}$CN (1-0), H$^{13}$CO$^{+}$ (1-0) HN$^{13}$C (1-0), H$^{15}$NC (1-0) and HC$^{15}$N (1-0) are integrated using the FWHM of HCO$^{+}$ (1-0) for each pixel. For the other lines, their moment 0 maps are integrated using their own FWHM for each pixel.

\subsection{Distributions of spectral line}

In Figure \ref{fig:Halpha}, we present H${\alpha}$, H41$\alpha$, and 3mm continuum. The optical image is observed with the Hubble Space Telescope Advanced Camera for Surveys (HST/ACS), taken from \cite{Mutchler2007}. Figure \ref{fig:HC3N_H41a_HCN_continuum} shows the integrated intensity maps of H$^{13}$CN (1-0), HCN (1-0), HC$_3$N (10-9), and H41$\alpha$ line emissions, which are detected in the central starburst disk. According to the H$^{13}$CN (1-0) emission peaks, we define four positions: E1, E2, W1, and W2, which are marked in the top left panel of Figure \ref{fig:HC3N_H41a_HCN_continuum}.

All lines roughly follow the CO brightness distribution with the triple-peaked structure along the major axis of M 82 \citep{Walter2002}. HCN (1-0) distribution is slightly more extended than H$^{13}$CN (1-0), HC$_3$N (10-9), and H41$\alpha$, which are concentrated in the galactic plane. The distribution and kinematics of our molecular emission in our results are consistent with the previous studies
\cite[HCO; H92${\rm \alpha}$; H53${\rm \alpha}$; CO (2-1); CN (1-0); N$_2$H$^+$ (1-0); CO (1-0);][]{Garcia2002,Rodriguez2004,Ginard2015,Krieger2021}. 


Here, we focus on the innermost region, especially for the three peak regions identified. HCN (1-0) appears stronger towards the outer positions (E1 and W2) than in the inner positions (E2 and W1). However, the optically thin H$^{13}$CN (1-0), H$^{13}$CO$^{+}$ (1-0) (see Figure \ref{fig:isotopicline}) and HC$_3$N (10-9) reveal the spatial distribution of dense gas is asymmetric along the major axis of M 82, with the northeast lobe (NE lobe) more intense than those of southwest lobe (SW lobe). More young massive stars or star clusters are located at the SW lobe (right), traced by H41$\alpha$ and 3mm continuum emissions.

\subsubsection{HC$_3$N (10-9), HCN (1-0), H41$\alpha$, and 3mm continuum}

HC$_3$N (10-9), as a tracer of warm and dense gas \citep{Tanaka2018}, is concentrated in the triple-peaked regions without extended emission. The strongest HC$_3$N emission peak is located at the NE lobe (left). The other two emission peaks at the SW lobe show similar integrated intensities, which are $\sim$ 50 percent of the brightest peak. These results indicate that the bulk of dense gas accumulates at the NE lobe. Our data better resolves the fainter emission of HC$_3$N (10-9) and shows more extended morphology with a higher sensitivity than the detection of \cite{Chidiac2020}. The line profiles of HC$_3$N (10-9) are very similar to the detections with IRAM 30m telescope in the corresponding positions \citep{Aladro2011b,Li2022}.

HCN (1-0) emission is very bright along the galactic plane and reveals very strong extended emission at least 10\arcsec  above or below the central starburst, which is slightly more extended than the other lines. The HCN (1-0) emission in the molecular disk could be optically thick, especially in the triple-peaked regions, due to the amount of dense molecular gas that accumulates in the central molecular disk indicated by HC$_3$N (10-9) emission.


H41$\alpha$ recombination line is a tracer of ionized gas. 
Massive stars emit strong UV radiation and stellar winds to form HII regions, which produce recombination lines such as H41$\alpha$. Therefore, H41$\alpha$ emission coincides with regions of recent massive star formation in M 82. We find a brighter and more extended H41$\alpha$ emission at the SW lobe compared to the NE lobe, in agreement with the distributions of H92${\rm \alpha}$ and H53${\rm \alpha}$ from \citep{Rodriguez2004}. On average, the intensity of H41$\alpha$ at the SW lobe is approximately twice that at the NE lobe. The distributions of the ionized gases indicate that there are more young massive stars or star clusters located at the SW lobe, which is dominated by more active starburst activity. However, the optically thin dense gas tracers of HC$_3$N (10-9), H$^{13}$CN (1-0), and H$^{13}$CO$^{+}$ (1-0) (Sec. \ref{isotopic lines}) are brighter at the NE lobe than those of SW lobe.


The 3mm continuum emission is fainter at the NE lobe compared with its counterparts at the SW lobe, in good agreement with the distributions of other radio continuum emissions at 1.5, 5, 15, and 22 GHz  \citep{Carlstrom1991} and 8.3 and 43 GHz \citep{Rodriguez2004}.
Previous studies of \cite{Puxley1989}, \cite{Carlstrom1991}, and \cite{Salas2014} have suggested that 3mm continuum emission can trace the site of recent star formation \citep{Salas2014}, since it is dominated by the optically thin free-free emission in the central region of M 82. Hence, the distribution of 3mm continuum emission indicates that the young massive stars are mostly located at the SW lobe. Furthermore, the brightness temperature of 3mm continuum emission is about 1.0 K at the emission peak, and it follows the same spatial distribution of H41$\alpha$.

\subsubsection{Rare HCN, HCO$^+$ and HNC isotopologues} \label{isotopic lines}

\begin{figure*}
	\includegraphics[scale=0.50]{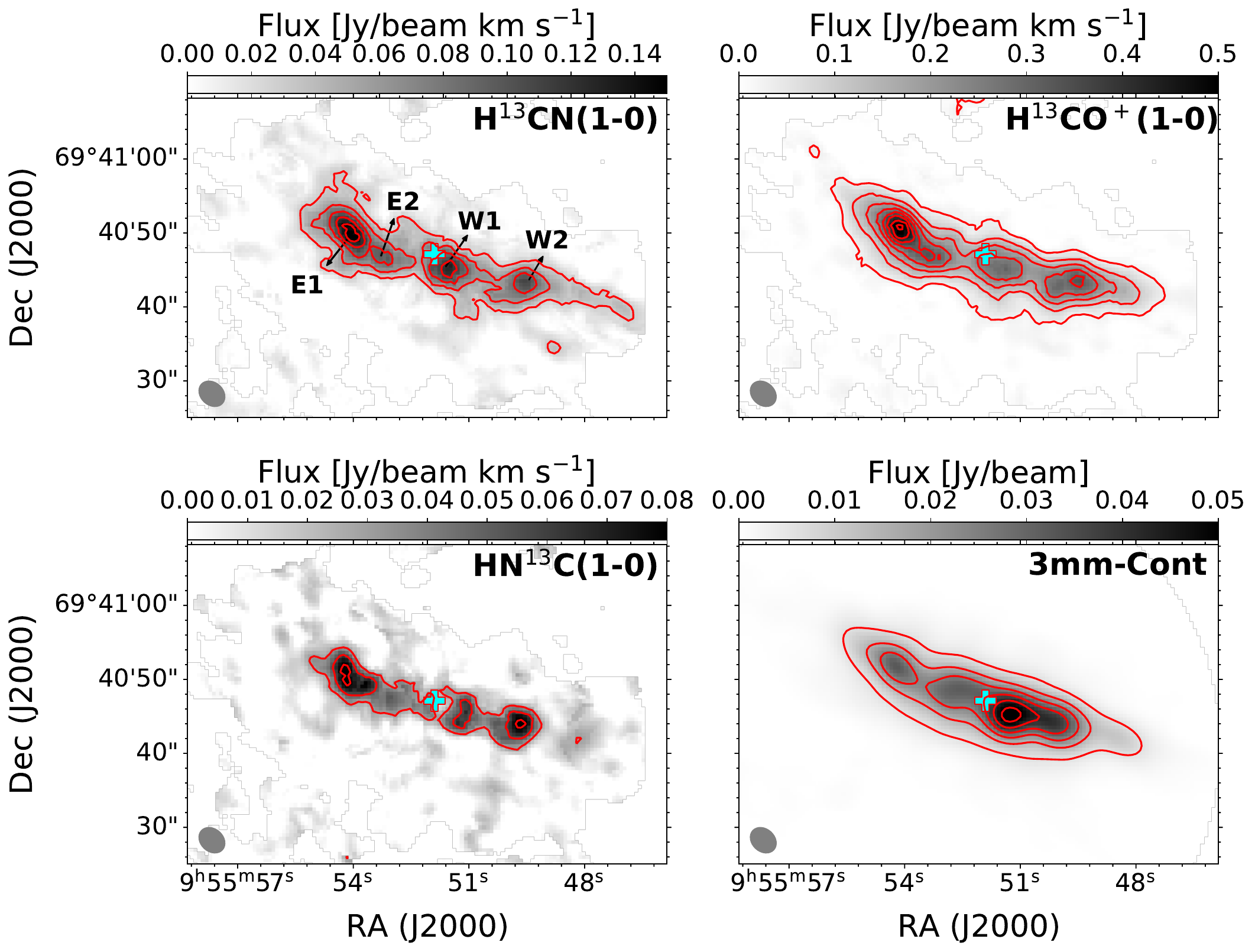}
    \caption{Line-integrated intensity maps of the H$^{13}$CN, H$^{13}$CO$^{+}$, HN$^{13}$C (1-0) and 3mm continuum. The synthesized beam is shown at the bottom left corner of the panels as a shaded ellipse. The dynamical center of the galaxy is marked by the central filled plus. Contour levels are 0.03 Jy beam $^{-1}$ km s$^{-1}$to 0.17 Jy beam $^{-1}$ km s$^{-1}$in steps of 0.035 Jy beam $^{-1}$ km s$^{-1}$for H$^{13}$CN, 0.03 Jy beam $^{-1}$ km s$^{-1}$to 0.52 Jy beam $^{-1}$ km s$^{-1}$in steps of 0.07 Jy beam $^{-1}$ km s$^{-1}$for H$^{13}$CO$^+$, 0.03 Jy beam $^{-1}$ km s$^{-1}$to 0.09 Jy beam $^{-1}$ km s$^{-1}$in steps of 0.03 Jy beam $^{-1}$ km s$^{-1}$for HN$^{13}$C. The contour levels of 3mm continuum are similar to Figure \ref{fig:Halpha}. First contour is at 3 $\times \sigma$ level.}
    \label{fig:isotopicline}
\end{figure*}

H$^{13}$CN (1-0), HN$^{13}$C (1-0), HC$^{15}$N (1-0) and H$^{15}$NC (1-0) are first presented at a giant molecular clouds (GMCs) scale resolution ($\sim$ 3.8\arcsec\ = 65 pc) in M 82. Figure \ref{fig:isotopicline} shows the moment 0 maps of H$^{13}$CN (1-0), H$^{13}$CO$^{+}$ (1-0), and HN$^{13}$C (1-0) and the $^{15}$N-bearing isotopologues are described in Appendix \ref{sec:$^{15}$N-bearing}. Their morphology resembles to the spatial distributions of main species, and they are more concentrated in the molecular disk. They are detected with signal-to-noise ratio $>$10 in the brightest regions. Similar to the distribution of HC$_3$N (10-9), the strongest emission peak of H$^{13}$CN (1-0), H$^{13}$CO$^{+}$ (1-0) and HN$^{13}$C (1-0) are located at E1 position. Furthermore, H$^{13}$CO$^{+}$ (1-0) and HN$^{13}$C (1-0) distributions behave similarly, with brighter emissions in the outer part than those of the inner part.  This asymmetry has also been found in ionized gas and several molecular line emissions, such as CN (1-0) and N$_2$H$^+$ (1-0) \citep{Achtermann1995,Rodriguez2004,Ginard2015}. 

\begin{figure}
	\includegraphics[scale=0.40]{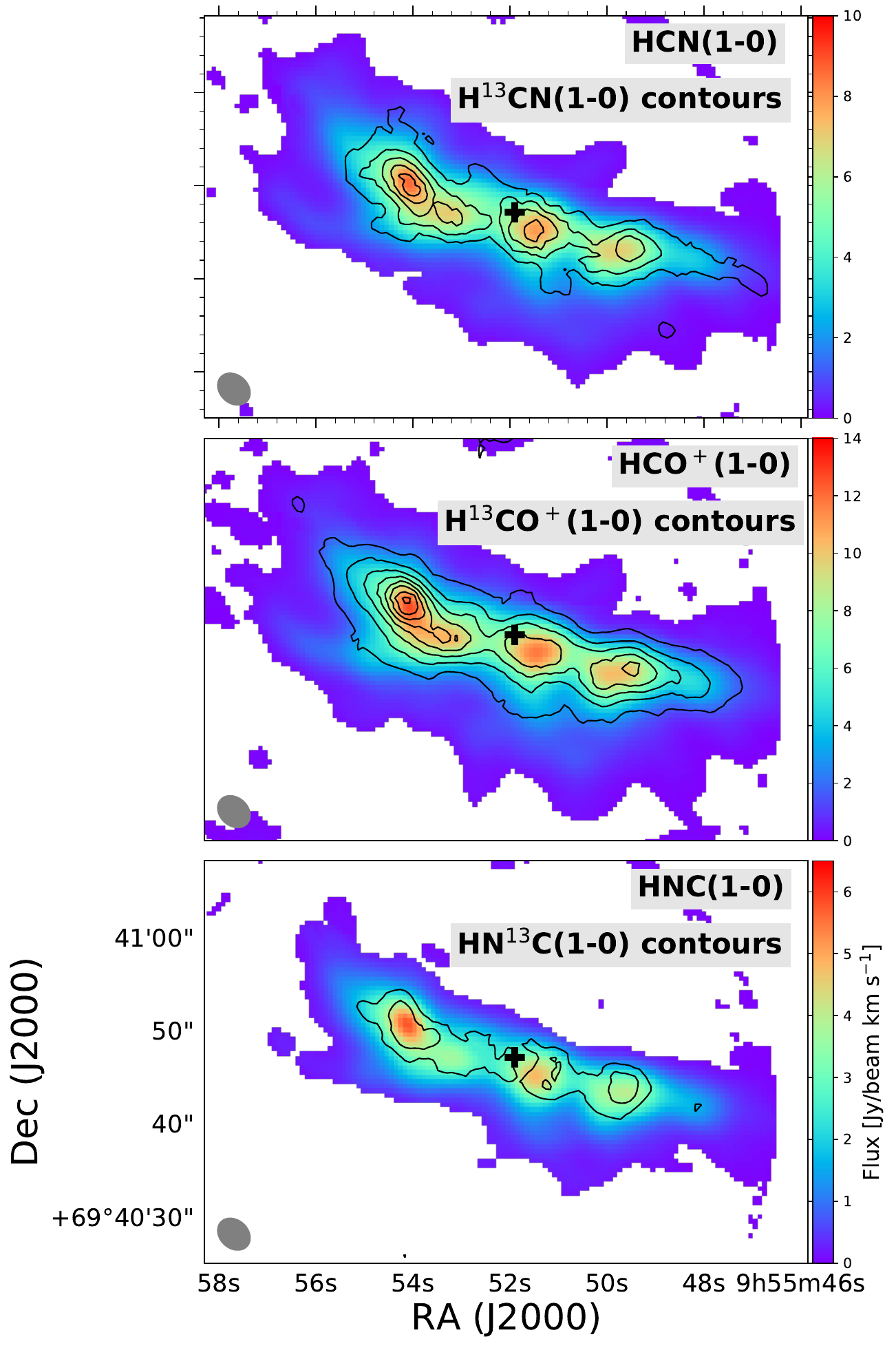}
    \caption{Line-integrated intensity maps of HCN, HCO$^+$ and HNC, which are ovelaid with H$^{13}$CN (1-0), H$^{13}$CO$^{+}$ (1-0) HN$^{13}$C (1-0) contours, respectively. The beam is shown at the bottom left corner of the panels as a shaded ellipse. The dynamical center of the galaxy is marked by the central filled plus. The contours levels are same as  Figure \ref{fig:isotopicline}.}
    \label{fig:main_isotopologue}
\end{figure}

\begin{figure*}

\includegraphics[scale=0.27]{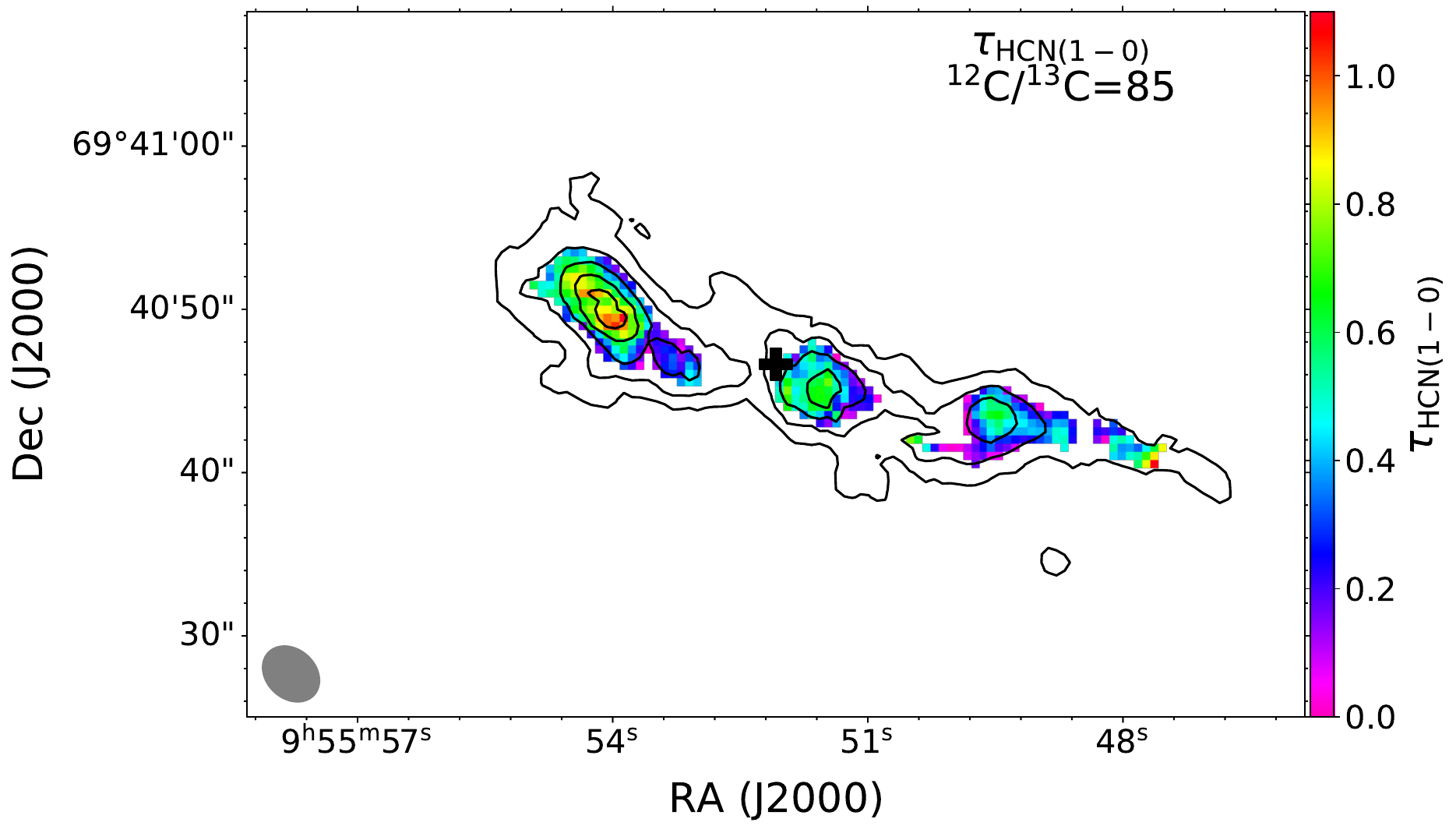}
\includegraphics[scale=0.27]{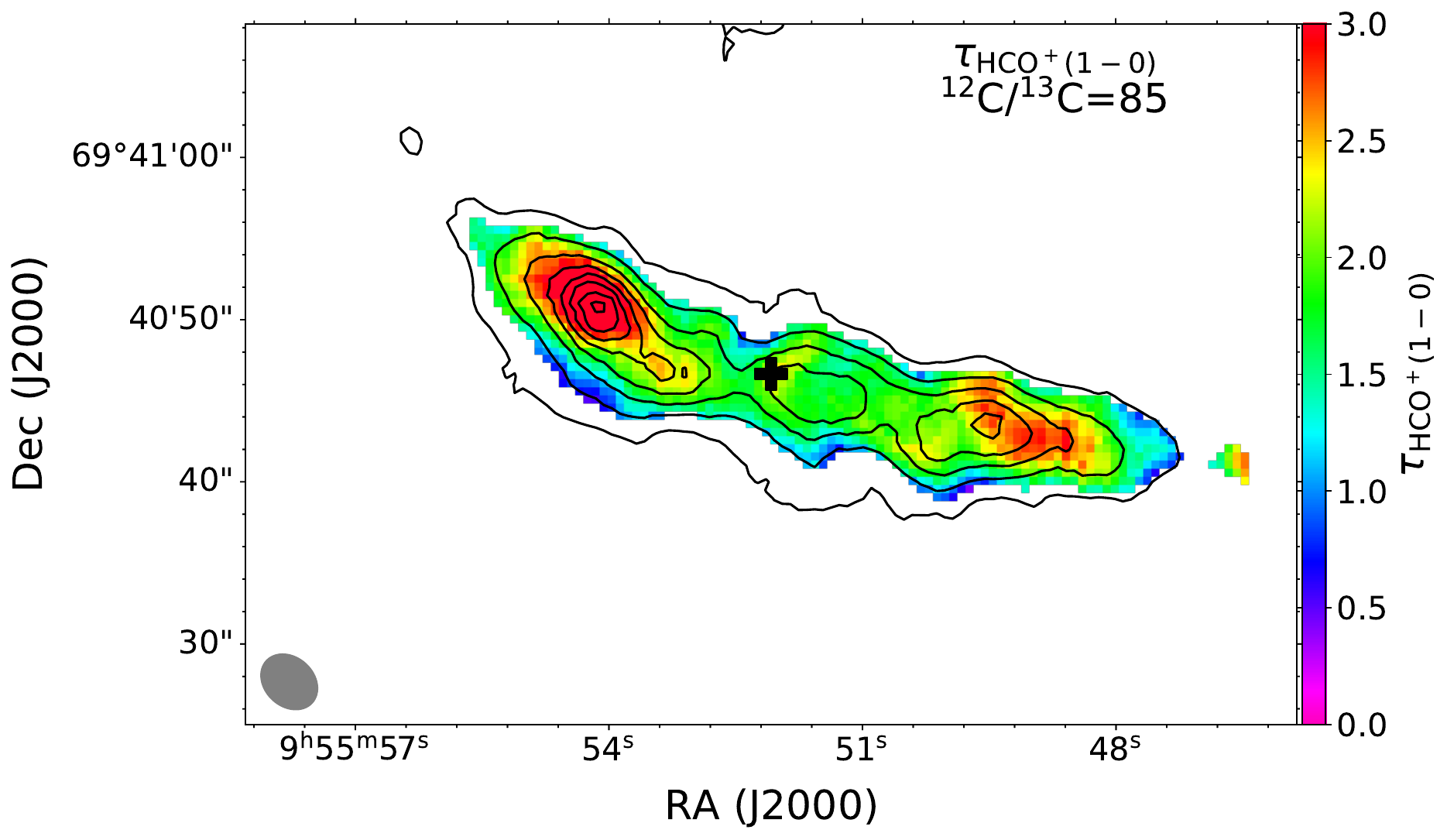}
\includegraphics[scale=0.27]{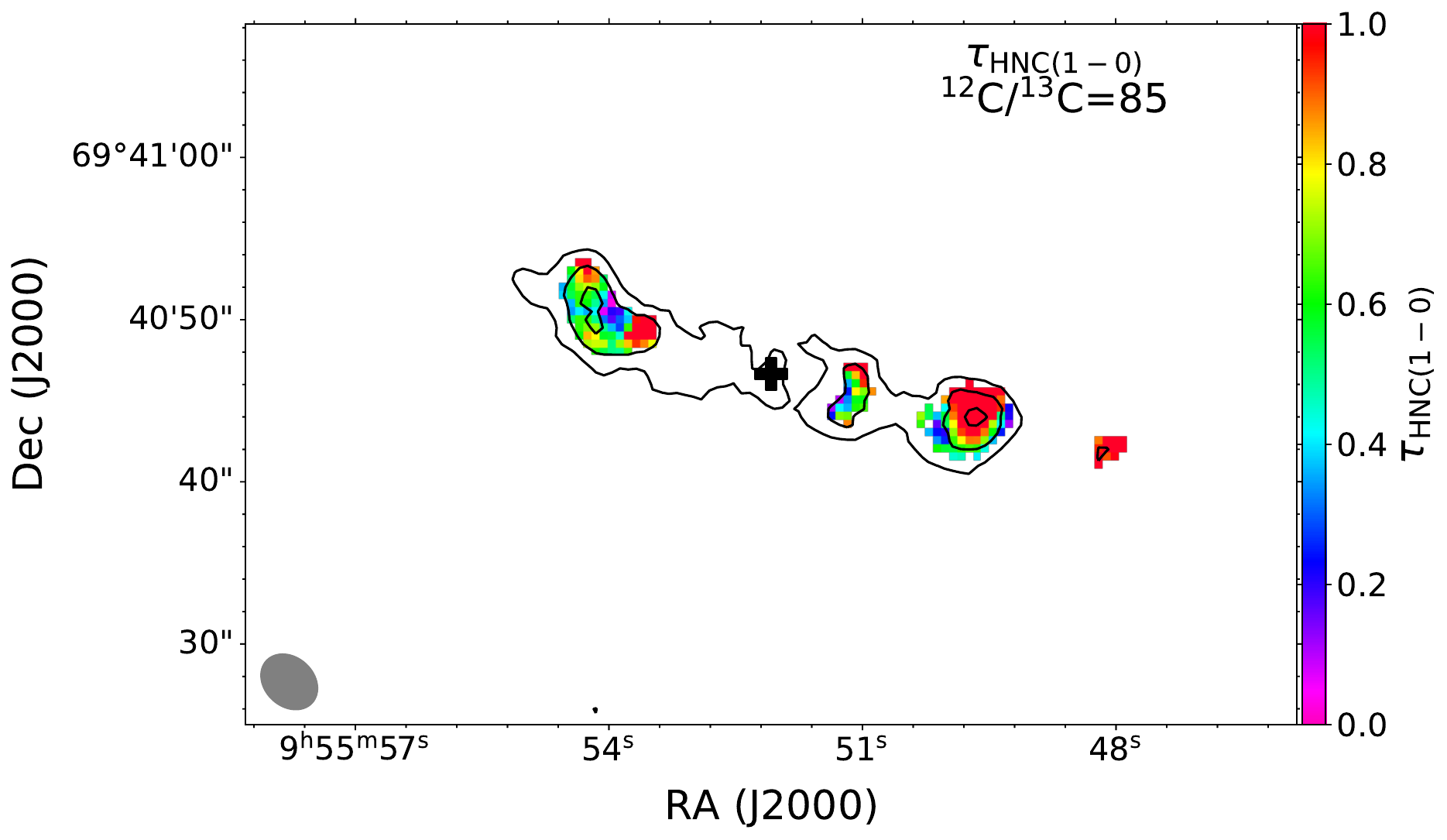}
\caption{The optical depth of HCN (1-0), HCO$^+$ (1-0) and HNC (1-0), which are ovelaid with H$^{13}$CN (1-0), H$^{13}$CO$^{+}$ (1-0) HN$^{13}$C (1-0) contours, respectively. The dynamical center of the galaxy is marked by the central filled plus. The contours levels are same as Figure \ref{fig:isotopicline}.}
\label{fig:optical_depth}
\end{figure*}

\section{Discussion}\label{sec:discussion}

\subsection{The star formation activity across the central starburst disk}

\begin{table*}
\centering

\caption{\textbf{The line flux in four H$^{13}$CN-bright regions}}
\label{tab:flux}
\begin{tabular}{lllll}
\hline
\hline
             
{Line}              & {E1}       & {E2}    & {W1}    & {W2} \\
\hline       
HC$_3$N (10-9)      & 110.6$\pm$0.3   &54.0$\pm$ 0.1    & 76.1$\pm$0.2   & 76.7$\pm$0.2 \\
H$^{13}$CN (1-0)    & 14.9$\pm$0.04     &10.1$\pm$0.02     & 12.6$\pm$ 0.03  &13.3$\pm$0.03\\        
H41${\alpha}$     &65.8$\pm$ 0.1      & 46.2$\pm$ 0.2   & 140.6$\pm$0.4  &95.2$\pm$0.4\\
3mm continuum      & 3.1$\pm$ 0.006      & 2.7$\pm$ 0.007    & 5.3$\pm$ 0.01   &2.7$\pm$0.009\\ 
\hline

\end{tabular}\\

\begin{minipage}{1.3\columnwidth}
 \vspace{1mm}
{\bf Notes:}  The flux is measured within a beam-sized region centered on the four intensity peaks of H$^{13}$CN. The units are Jy beam$^{-1}$ km s$^{-1}$ and Jy beam$^{-1}$ for molecular lines and 3mm continuum, respectively. The uncertainty is the standard deviation.
\end{minipage}

\end{table*}


Based on the images of H41$\alpha$, 3mm continuum, HC$_3$N (10-9), and H$^{13}$CN (1-0), we conducted an in-depth study of the relationship between star formation and the dense molecular clouds. H41$\alpha$, 3mm continuum present intense emission towards the SW lobe, which is the site of recent star formation. Their fluxes at the SW lobe are higher by a factor of $\sim$1.7-2.1 than those of the NE lobe (see Table \ref{tab:flux}). However, the distributions of HC$_3$N (10-9) and H$^{13}$CN (1-0) show a different trend, which suggests a factor of $\sim$1.4 lower peak fluxes at the SW lobe relative to the NE lobe.


The distribution of H41$\alpha$ emissions strongly suggests that starburst activities are different at the NE lobe and SW lobe. This is consistent with the findings from mid-infrared (mid-IR) that the SW lobe contains the two most active star-forming regions (W1 and W2 in Figure \ref{fig:HC3N_H41a_HCN_continuum}) \citep{Lipscy2004,Matsushita2005}. 
On the other hand, we compare the distributions of H41$\alpha$ , H$\alpha$, and 3 mm continuum emission, in Figure \ref{fig:Halpha}. The peaks of H41$\alpha$ and 3mm continuum emission correspond to the high dust opacity region from the H$\alpha$ image. These results indicate that massive stars have already formed in the clouds at the SW lobe. Furthermore, the distributions of dense gas tracers suggest that a large concentration of dense molecular gas exists at the NE lobe.



For the SW lobe, the strong free-free radio emission and X-ray emissions show evidence that ionized gas could be sufficient to drive the compression of dense molecular clouds to trigger star formation \citep{Keto2005}.  Based on the H41$\alpha$ emission, it is expected that the UV radiation from starburst is much stronger at the SW lobe. 

For the NE side, the star formation is relatively quiescent, where dense molecular clouds are found to be in expansion \citep{Keto2005}. The strong HC$_3$N (10-9) and H$^{13}$CN (1-0) emissions prove that the molecular gas reservoir to form new stars is not exhausted, suggesting that these regions might be sites of future star formation.

\subsection{The dense gas properties traced by HCN (1-0), $^{13}$C-bearing species and HC$_3$N (10-9)}

As shown in Figure \ref{fig:main_isotopologue}, HCN, HCO$^+$, and HNC (1-0) show similar spatial distribution in the nuclear starburst of M 82. The triple peak intensities do not show a significant difference. Since HCN, HCO$^+$, and HNC (1-0) are usually optically thick, thus, their flux cannot accurately trace the dense gas mass. However, their $^{13}$C isotopologue counterparts, namely H$^{13}$CN, H$^{13}$CO$^{+}$, and HN$^{13}$C (1-0), could better probe the actual dense gas structure. 
Their distributions reveal that dense gas emission is asymmetric in these two lobes, with the NE lobe more intense than that of the SW lobe. The isotopologues show that the brightest emission is from the E1 position, and a relatively weak emission is from the other three positions.
Our results are consistent with the existence of a gas density gradient, which increases from the SW lobe to the NE lobe \citep{Petitpas2000}.

\begin{figure}
	\includegraphics[scale=0.55]{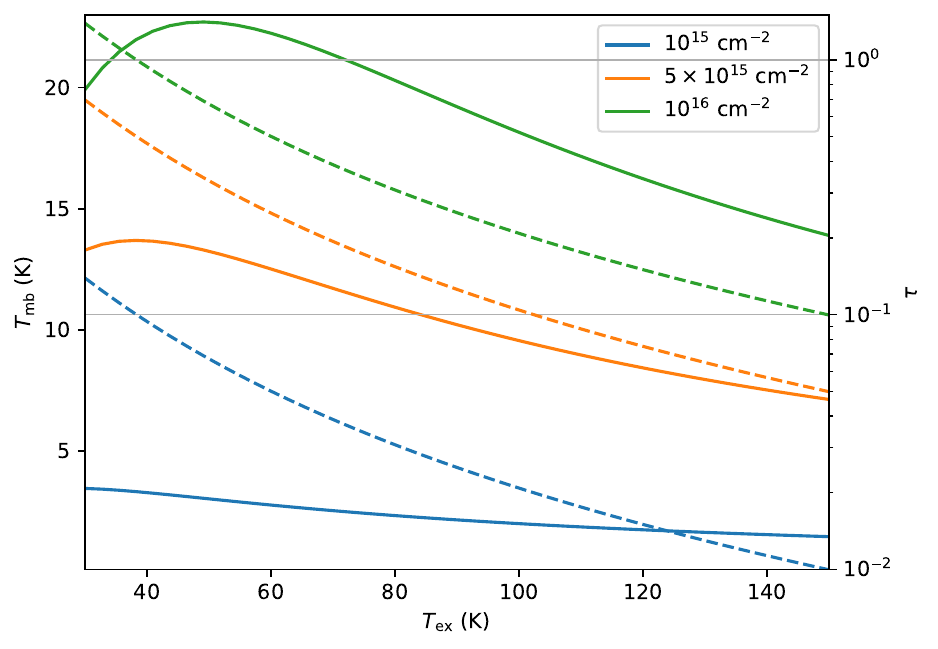}
    \caption{Assuming LTE, we calculate intensity integrate and optical depth of HC$_3$N (10-9) vary with the gas excitation temperature. The orange, green, and blue curves represent the integrate intensity. The dashed curves represent the optical depth.}
    \label{fig:HC3N_Temperature}
\end{figure}

HC$_3$N (10-9) line shows a similar behavior as the above $^{13}$C-bearing molecules (see Figure \ref{fig:HC3N_H41a_HCN_continuum}). The stronger HC$_3$N emission in the NE lobe may be related to two factors: (1) more dense gas accumulation (higher column density) and (2) differences in excitation conditions. The former is expected since the stronger UV radiation or cosmic-rays (CRs) generated by massive stars at the SW lobe would efficiently destroy HC$_3$N \citep{Costagliola2010}. On the other hand, since the gas properties (e.g., temperature, density) could be different in the two regions, the excitation temperature could be different. Supposing that the two regions have similar gas column density, an increase of gas excitation temperature by a factor of 5 could lead to a decrease of line intensity and optical depth by a factor of $\sim$2, assuming local thermodynamic equilibrium (LTE) and uniform abundance (see Figure \ref{fig:HC3N_Temperature}). Without multiple transitions to constrain the gas physical properties, we cannot make a certain conclusion.


\subsection{Optical depths of dense gas tracers}
\subsubsection{The distribution of optical depth}
\begin{table}
\centering

      \caption{\textbf{$\rm ^{12}C$/$\rm ^{13}C$ measured in external galaxies.}}
      \label{tab:12C_13C}
      
\begin{tabular}{llll}
\hline
\hline
             
{Galaxy}   & {type} & {$\rm ^{12}C$/$\rm ^{13}C$}  & {Ref.} \\
\hline       
M82        & starburst & $>$40   & 1,2\\
           &           & $>$130  & 3\\
\hline           
NGC253     &starburst & $\sim$40   & 1,2\\
           &           & $\sim$80  & 3\\ 
\hline
NGC1614    &LIRG       &$\sim$130  &4\\
\hline
Mrk231     &ULIRG      &$\sim$100  &2\\
\hline
Arp220     &ULIRG      &$\sim$100  &5\\
\hline
Arp193     &ULIRG      &$\sim$150  &6\\
\hline

\end{tabular}\\
        
\begin{minipage}{0.95\columnwidth}
 \vspace{1mm}
{\bf Notes:} (1) \cite{Henkel1998}; (2) \cite{Henkel2014}; (3) \cite{Martin2010}; (4) \cite{Sliwa2014}; (5) \cite{Gonzalez2012}; (6) \cite{Papadopoulos2014}.
\end{minipage}

\end{table}

The $\rm ^{12}C$/$\rm ^{13}C$ ratio typically ranges from 40 to 130 across different types of galaxies (see Table~\ref{tab:12C_13C}).
Here, we adopt \textbf{a median} isotopic abundances ratio of $^{12}$C/$^{13}$C=85 to calculate the optical depth in central regions of M82, which shows similar properties as ULIRGs. More details on the discussion about the variation of $\rm ^{12}C$/$\rm ^{13}C$ can be found in \cite{Li2022}. 
Assuming that the range of $\rm ^{12}C$/$\rm ^{13}C$ is uniformly distributed between 40 and 130, the standard deviation of 26 is obtained by dividing the range (130-40 = 90) by $\sqrt{12}$. 
This corresponds to 26/85 $\approx$ 31\% relative to the adopted  ratio of $^{12}$C/$^{13}$C=85.
This results in 1 $\sigma$ uncertainty in optical depth of 31\%.

We obtain the first spatial distributions of HCN, HCO$^+$, and HNC (1-0) optical depth, which are presented in Figure \ref{fig:optical_depth}. The detailed calculation of the optical depth is described in our previous work of \cite{Li2022}. On average, the optical depths of HCN (1-0) and HNC (1-0) are about twice smaller than HCO$^+$ (1-0). Their optical depths show a slight variation along the major axis of the continuum disk, which is consistent with $\sim$ a few hundred parsec observations \citep{Li2022}. Compared with the optical depths of HCN (1-0) and HCO$^+$ (1-0) in these four distinct positions, the highest optical depth is found at the E1 position, which might result from amounts of dense molecular gas accumulating at the E1 position. While, for the other three positions, the star formation activities and subsequent high rate of supernova explosions could consume or expel the gas via the starburst wind, resulting in the low optical depths of HCN (1-0) and HCO$^+$ (1-0). HNC (1-0) and HCO$^+$ (1-0) show two high opacity spots associate with E1 and W2 clouds and uniform elsewhere.


%


\begin{table*}
\centering

      \caption{\textbf{The isotopic line ratio measured in four H$^{13}$CN-bright beam-sized regions.}}
      \label{tab:line ratio}
      
\begin{tabular}{llllllllllll}
\hline
\hline

		    &	E1			    &	E2		     	&	W1			&	W2			\\
\hline
\bf{HCN/H$^{13}$CN}		&				&				&				&				\\
Range		&	52.0	$-$	79.3	&	59.6	$-$	86.5	&	59.9	$-$	89.2	&	62.5	$-$	105.2	\\
Average		&	65.1	$\pm$	9.1	&	74.6	$\pm$	10.4	&	69.1	$\pm$	9.7	&	76.6	$\pm$	10.7	\\
\hline		
\bf{HCO$^{+}$/H$^{13}$CO$^{+}$}		&				&				&				&				\\
Range		&	22.5	$-$	48.5	&	28.6	$-$	54.2	&	32.8	$-$	52.1	&	28.3	$-$	48.9	\\
Average		&	30.1	$\pm$	4.2	&	37.0	$\pm$	5.2	&	40.0	$\pm$	5.6	&	33.2	$\pm$	4.6	\\
\hline		
\bf{HNC/HN$^{13}$C}		&				&				&				&				\\
Range		&	44.7	$-$	83.6	&	44.7	$-$	67.2	&	47.1	$-$	81.0	&	37.4	$-$	89.2	\\
Average		&	63.8	$\pm$	8.9	&	51.4	$\pm$	7.2	&	64.0	$\pm$	9.0	&	56.6	$\pm$	7.9	\\
\hline		

\end{tabular}\\

\end{table*}

\begin{figure*}

\includegraphics[scale=0.45]{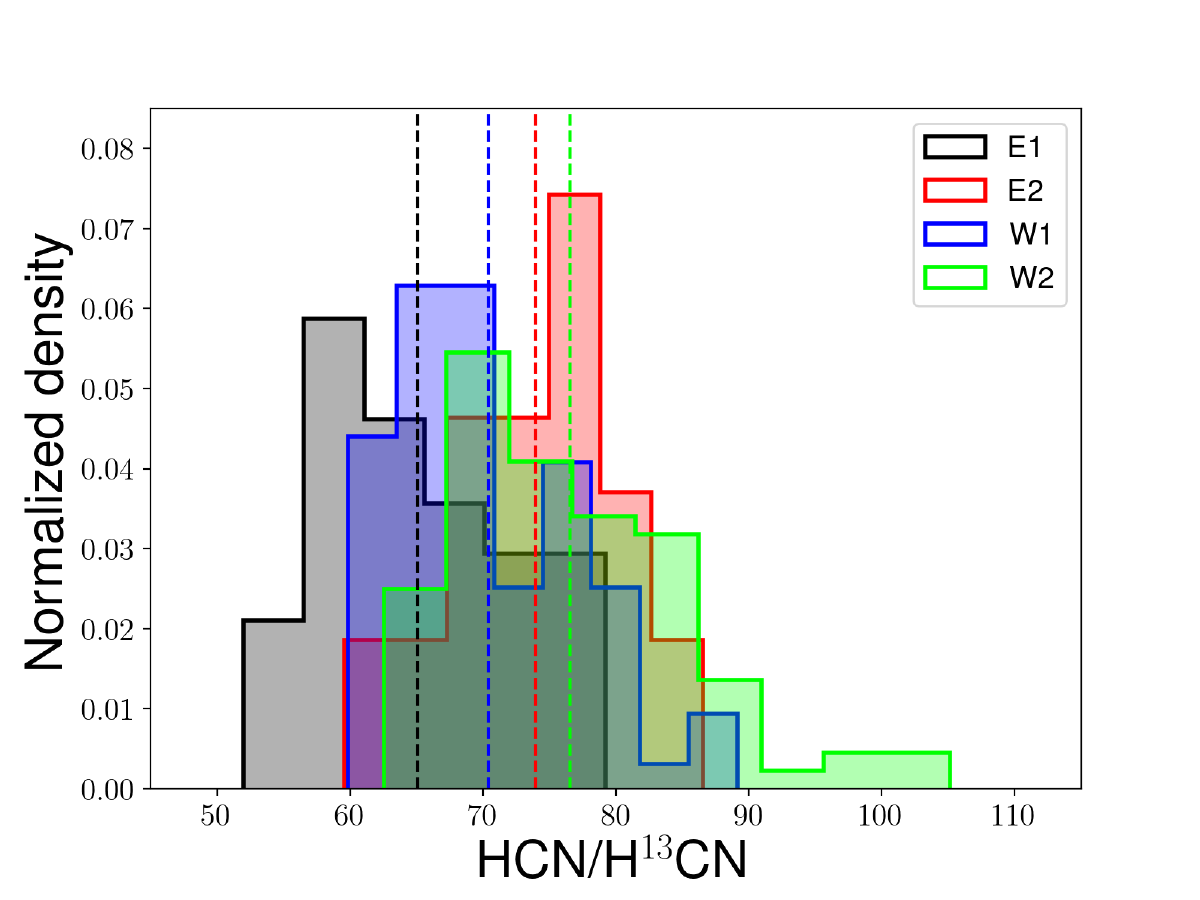}
\includegraphics[scale=0.45]{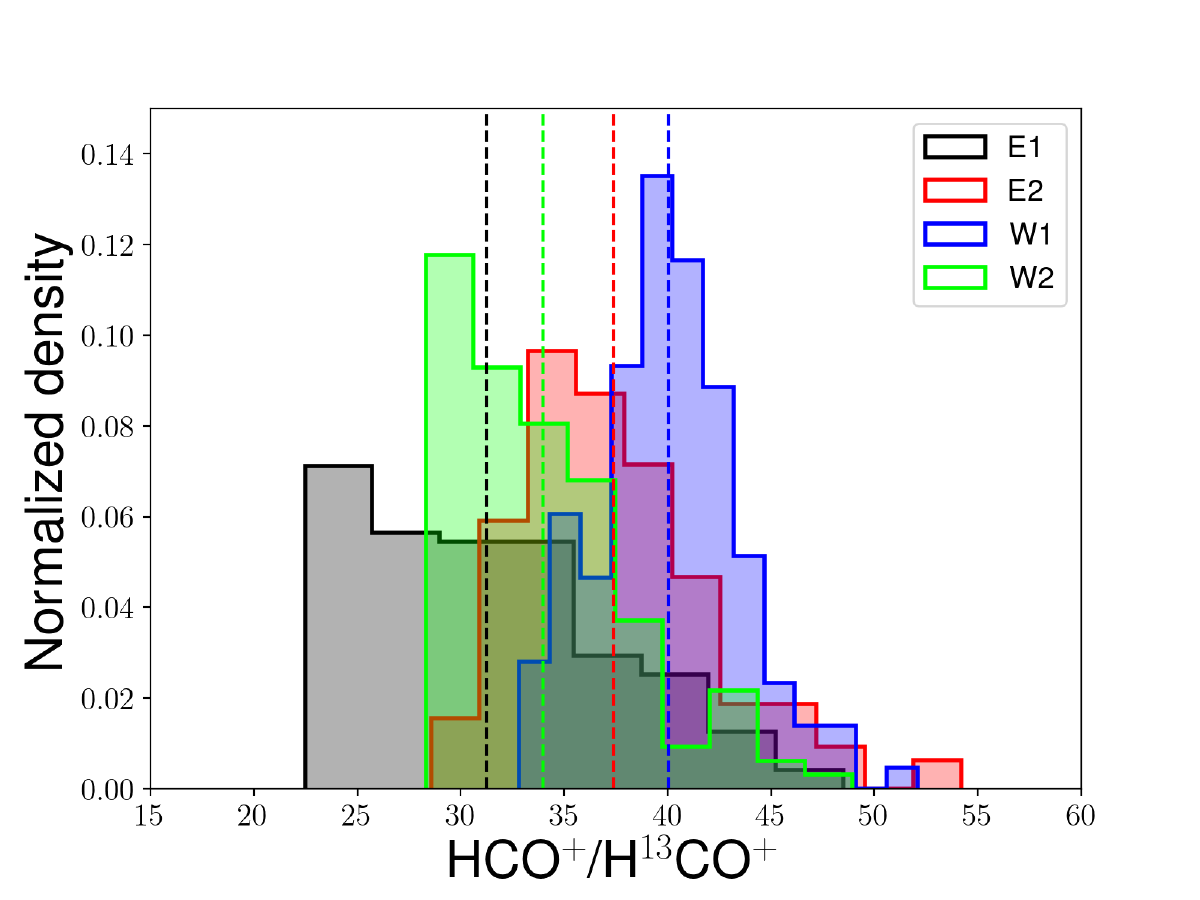}
\includegraphics[scale=0.45]{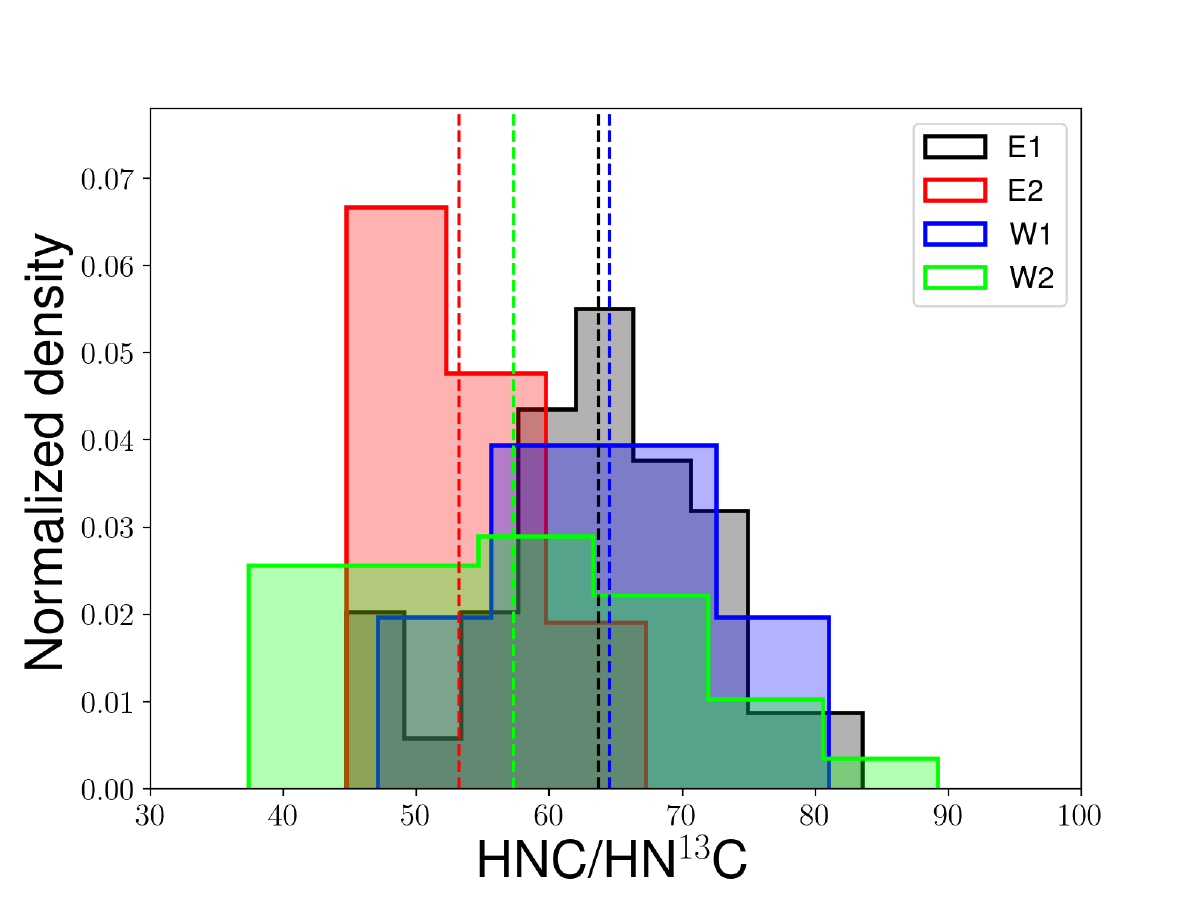}
\caption{Gaussian kernel density estimate of isotopic line ratio distribution. Top left: HCN/H$^{13}$CN. Top right: HCO$^+$/H$^{13}$CO$^{+}$. Bottom: HNC/HN$^{13}$C. The vertical dashed-dotted lines represent the mean value of the line ratio.}
\label{fig:histogram}
\end{figure*}

\subsubsection{Variation of $^{12}$C/$^{13}$C abundance ratios}

Basing on our previous study of \cite{Li2022}, $^{12}$C/$^{13}$C abundance ratios should
increase from center to outskirt along the major axis of M82. $^{12}$C/$^{13}$C abundance ratios could represent the star-formation 
history \citep{Wilson1994, Henkel2010}. However, the observed $^{12}$C/$^{13}$C ratios are not only from the current starbursts contribution, but also produced from the all starbursts in history.
Besides, Galactic chemical evolution models predict that the $^{12}$C/$^{13}$C ratio would vary by 
a factor of two, even if a strong starburst produce half of the stellar mass in a secular evolving 
galaxy \citep{Romano2017}.
 
On average, the isocopic ratios of HCN/H$^{13}$CN, HCO$^+$/H$^{13}$CO$^{+}$, and HNC/HN$^{13}$C are relatively constant with a variation $\sim$ 20-30\% in a beam-sized region
centered on the four intensity peaks of H$^{13}$CN (see Table \ref{tab:line ratio} and Figure \ref{fig:histogram}). Considering the possible radial gradient of the $^{12}$C/$^{13}$C abundance ratio, it is difficult to obtain accurate optical depths due to the lack of abundance measurements. 
However, our measured line ratios between the isotopologues can still set the lower limit for the abundance ratios.



\subsection{Origin of isotopic ratio variation}

\subsubsection{Physical conditions}
The average isotopic ratios at the NE lobe are lower than those of SW lobe. This could be attributed to differences in their optical depth. The HCN (1-0), HCO$^+$ (1-0) and HNC (1-0) emissions toward NE lobe could be optically thick while they could be optically thin toward SW lobe, due to amount of dense gas accumulates NE lobe. On the other hand, theoretically, we expect H$^{13}$CN (1-0), H$^{13}$CO$^{+}$ (1-0), and HN$^{13}$C (1-0) emissions to trace denser and cooler gas than HCN (1-0), HCO$^+$ (1-0) and HNC (1-0). Thus, the higher gas density would drive the low isotopic ratios at the NE lobe.

\subsubsection{Astrochemical effects}
\label{Appendix:astrochemical effects}
Selective photon dissociation prefers to destroy $^{13}$C and $^{15}$N bearing molecules \citep{Wilson1992,Savage2002}. This would increase the estimation of $^{12}$C/$^{13}$C abundance ratio in high UV fields. Previous studies suggest that PDR chemistry is propagating in the central region \citep{Garcia2002,Ginard2015}. CN (1-0) emission (PDR tracer) is highest toward W1, with the SW lobe more intense than the NE lobe \citep{Garcia2002,Chidiac2020}. The W1 position has higher isotopic ratios of  HCO$^+$/H$^{13}$CO$^{+}$, and HNC/HN$^{13}$C indicating that selective photo dissociation might play a key role.


Isotope fractionation could effectively reduce HCN/H$^{13}$CN, HCO$^+$/H$^{13}$CO$^{+}$, and HNC/HN$^{13}$C ratios at very low temperatures \citep{Smith1980,Woods2009,Szhucs2014,Colzi2020}. However, we suggest that fractionation effect does not play a key role in the variation of the line ratios, since the gas temperature is relatively high due to the starburst.

%

\subsection{Future prospect about the dense gas distribution in galaxies}
\label{sec:future}
Optically thin dense gas tracers would accurately trace gas structure or the distributions of gas density in star formation regions, compared with the main dense gas tracers. However, the isotopic lines of dense gas tracers are too faint to be detected in most galaxies. The HC$_3$N transitions, often found several times stronger than isotopologues of dense gas tracers, are very likely optically thin in most cases, and might be a better tracer to probe the spatial resolution of dense gas in galaxies. Studies of optically thin dense gas tracers are still rare in cases. ALMA with unprecedented sensitivity enables a systemic study, by observing a large sample of nearby galaxies.



\section{Conclusions}\label{sec:summary}

We present a NOEMA observation of HC$_3$N (10-9), H41$\alpha$, 3mm continuum emissions, HCN, HCO$^+$, HNC (1-0) and their isotopologues towards the central starburst region of M 82, with a spatial resolution 3.8\arcsec\ , $\sim$65 pc. The isotopologues H$^{13}$CN, HC$^{15}$N, HN$^{13}$C, H$^{15}$NC (1-0) are firstly obtained at GMC-scale resolution in M 82. We study the physical behavior of dense gas tracers as well as the spatial evolution of starburst activity. The main results can be summarized as follows.

(1) The overall morphology of all transitions are similar and distinct. Their distributions are dominated by the well-known triple-peaked structure (positions E1, W1 and W2).  These triple peak intensities of main dense gas tracers do not show a significant difference. However, optically thin HC$_3$N (10-9) and isotopologues H$^{13}$CN, H$^{13}$CO$^+$, HN$^{13}$C (1-0) line emissions are brighter at the NE lobe than those of SW lobe. Then, H41$\alpha$ shows a similar spatial distribution to 3mm continuum emissions, with the SW lobe more intense than the NE one. 



(2)Comparing with different dense gas tracers, dense gas distribution traced by optically thin HC$_3$N (10-9) and H$^{13}$CN (1-0) appear asymmetric, which suggest that a significant amount of dense gas accumulate at the NE lobe (left). HC$_3$N (10-9) and H$^{13}$CN (1-0) trace the interior of dense molecular clouds accurately comparing with main dense gas tracers, such as HCN, HCO$^+$, and HNC (1-0). 


(3) More active star formation regions are found at the SW lobe than those of the NE lobe, as traced by H41$\alpha$ and 3mm continuum emissions. Our results suggest that more massive star or star clusters have already been formed in the clouds at the SW lobe (right). The strong of HC$_3$N (10-9) and H$^{13}$CN (1-0) emissions prove that the molecular gas reservoir to form new stars are not exhausted at the NE lobe (left).

(4) On average, the isocopic ratios are relatively constant with a variation $\sim$ 20-30\% in a beam-sized region centered on the four intensity peaks of H$^{13}$CN. However, the W1 position has higher
HCO$^+$/H$^{13}$CO$^{+}$ and HNC/HN$^{13}$C ratios indicate that gas density and optical depth might play a key role for the different line ratios.



(5)The optical depths of HCN (1-0) and HNC (1-0) are around 1 at E1 positon, which are about three times smaller than HCO$^+$ (1-0),  adopting an isotopic abundance ratio of $^{12}$C/$^{13}$C=85. Considering the possible radial gradient of the $^{12}$C/$^{13}$C abundance ratio, it is difficult to obtain accurate optical depths due to the lack of abundance measurements. 
However, our measured line ratios between the isotopologues can still set the lower limit for the abundance ratios.


\begin{acknowledgments}

This work is supported by the National Natural Science Foundation of China grant (12103024,12041305, 12173016 and 12173067), and the fellowship of China Postdoctoral Science Foundation 2021M691531. We acknowledge the Program for Innovative Talents, Entrepreneur in Jiangsu. We acknowledge the science research grants from the China Manned Space Project with NO.CMS-CSST-2021-A08. 
This work is based on observations carried out under project number W20BT with the IRAM NOEMA Interferometer. IRAM is supported by INSU/CNRS (France), MPG (Germany) and IGN (Spain).
The HST data presented in this article were obtained from the Mikulski Archive for Space Telescopes (MAST) at the Space Telescope Science Institute. The specific observations analyzed can be accessed via \dataset[10.17909/snq6-h933]{https://doi.org/DOI}.

\end{acknowledgments}

%

%
%



\appendix
\section{Molecular line profile towards the triple-peak regions}
\begin{figure*}

\includegraphics[scale=0.27]{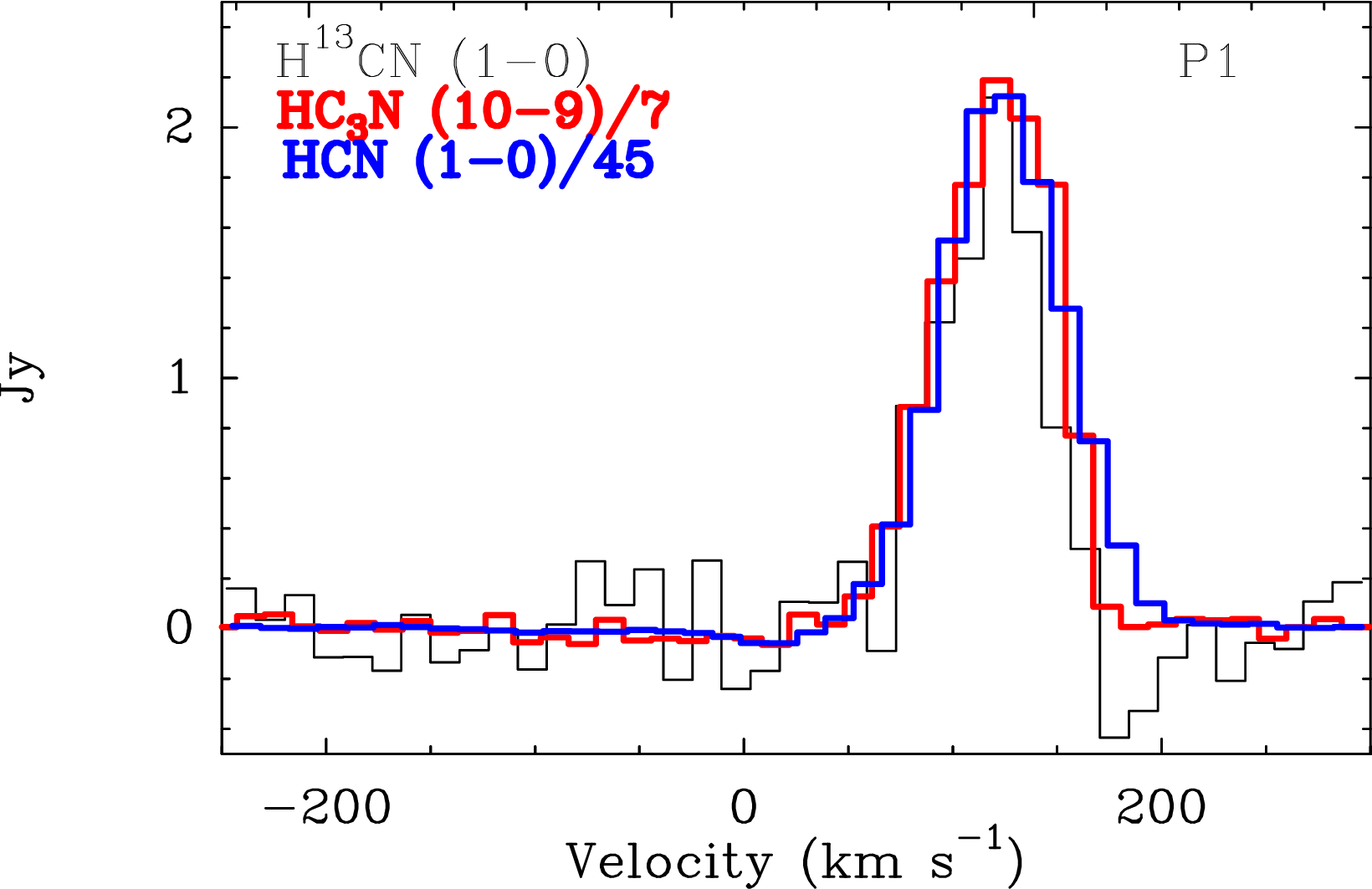}
\includegraphics[scale=0.27]{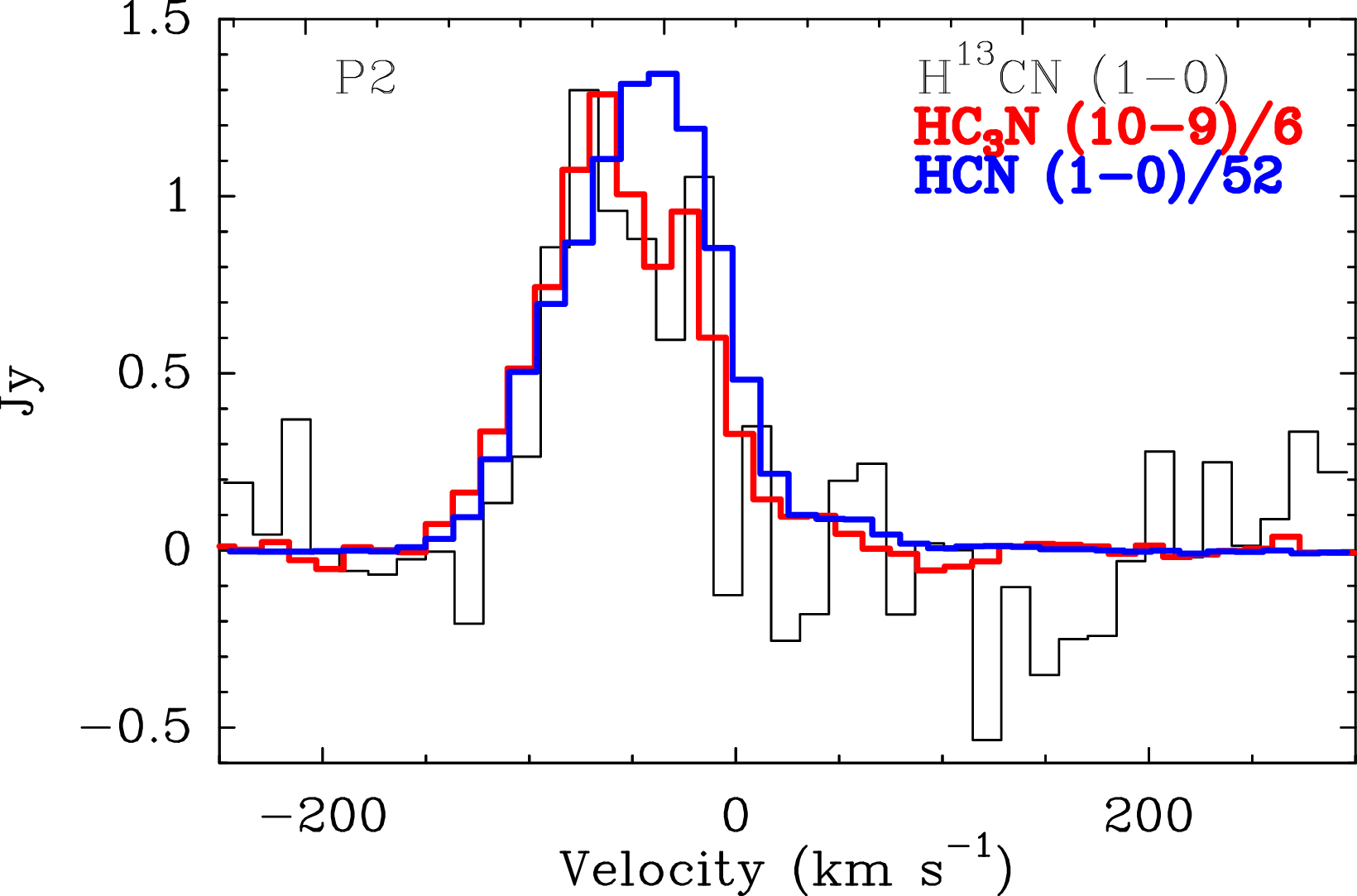}
\includegraphics[scale=0.27]{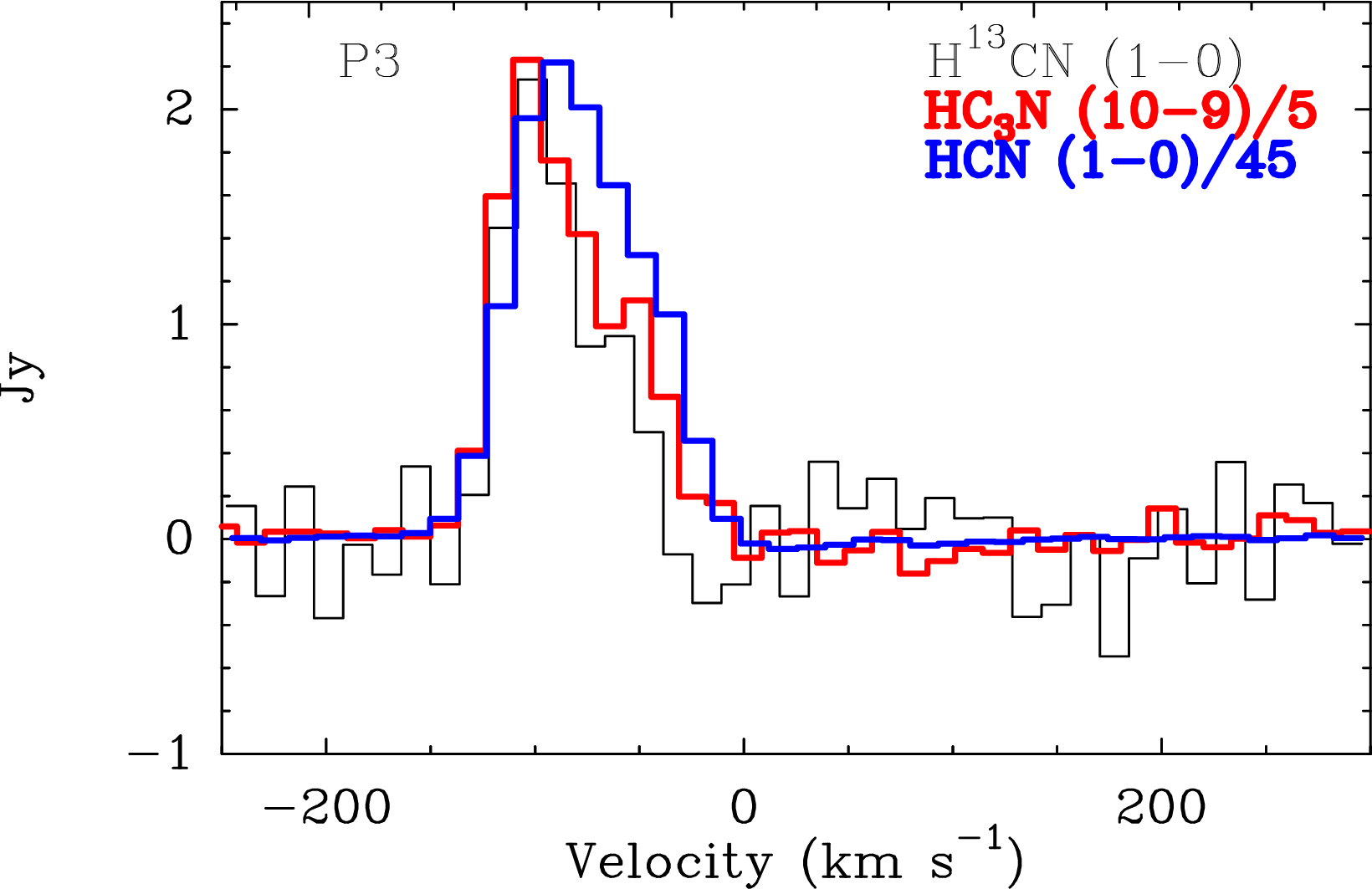}
\caption{Average spectra of one beam-sized regions (3.8 \arcsec\ $\times$ 3.8 \arcsec\ ) towards the triple-peak regions are shown for
HCN (1-0), HC$_3$N (10-9) and H$^{13}$CN (1-0). The velocity resolution of each spectra line is $\sim$15 km s$^{-1}$. }
\label{fig:spectralline}
\end{figure*}

The spectra lines of HCN (1-0), HC$_3$N (10-9) and H$^{13}$CN (1-0)
towards the triple-peak regions are shown in Figure \ref{fig:spectralline}, which are extracted from one beam size region (3.8 \arcsec\ $\times$ 3.8 \arcsec\ ). The velocity resolution of each spectra line is $\sim$15 km s$^{-1}$. On average, H$^{13}$CN (1-0) is $\sim$ 7 times fainter than HC$_3$N (10-9) and $\sim$ 45 times fainter than HCN (1-0). However, their line profiles are very similar and linewidths are around 120 km s$^{-1}$, which is consistent with $\sim$500 pc observation in M 82 \citep{Li2022}. It indicates that these three transitions spatially coexist
inside GMCs at $\sim$100 pc scales and they probably trace the cloud-to-cloud velocity dispersion or turbulence.

\begin{figure*}
	\includegraphics[scale=0.45]{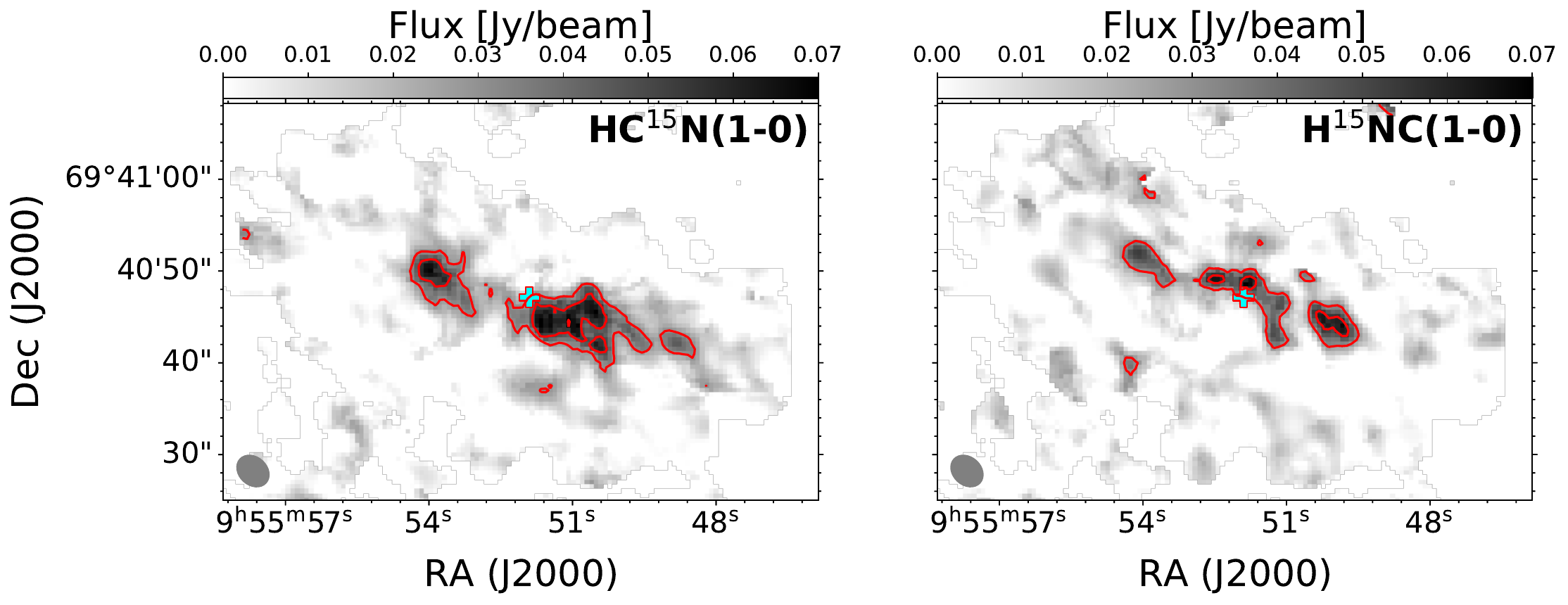}
    \caption{Line-integrated intensity maps of HC$^{15}$N (1-0) and H$^{15}$NC (1-0). The synthesized beam (3.8 \arcsec\ $\times$ 3.0 \arcsec\ ) is shown at the bottom left corner of the panels as a shaded ellipse. The dynamical center of the galaxy is marked by the central filled plus. No primary beam correction was applied to the images. The contour levels  are 0.03 Jy beam $^{-1}$ km s$^{-1}$to 0.07 Jy beam $^{-1}$ km s$^{-1}$by steps of 0.02 Jy beam $^{-1}$ km s$^{-1}$for HC$^{15}$N, 0.03 Jy beam $^{-1}$ km s$^{-1}$to 0.07 Jy beam $^{-1}$ km s$^{-1}$in steps of 0.025 Jy beam $^{-1}$ km s$^{-1}$for H$^{15}$NC.}
    \label{fig:isotopicline_N}
\end{figure*}

\section{The $^{15}$N-bearing molecular species}
\label{sec:$^{15}$N-bearing}
The emissions of HC$^{15}$N (1-0) and H$^{15}$NC (1-0) seem to be brighter in the inner part than those of the outer part, as shown in Figure \ref{fig:isotopicline_N}. Although their signal-to-noise levels are slightly lower than the other isotopologues, they follow the same spatial distribution in the major axis. HC$^{15}$N (1-0) might be blended with SO (2-1) with only a 200 km s$^{-1}$ offset in velocity, which cannot be separated. This line blending effect is most severe towards the inner part, which results in the strong HC$^{15}$N (1-0) emission there, as the results reported by \cite{Pineau1993,Aladro2011b} and \cite{Li2022}. For H$^{15}$NC (1-0), being the weakest molecular transition in our detection, was not detected by our previous single dish observations \citep{Li2022}. The lack of HC$^{15}$N (1-0) emission outside the region might be due to the low signal-to-noise ratio following the faint emission of HN$^{13}$C (1-0).

\begin{figure*}
\includegraphics[scale=0.27]{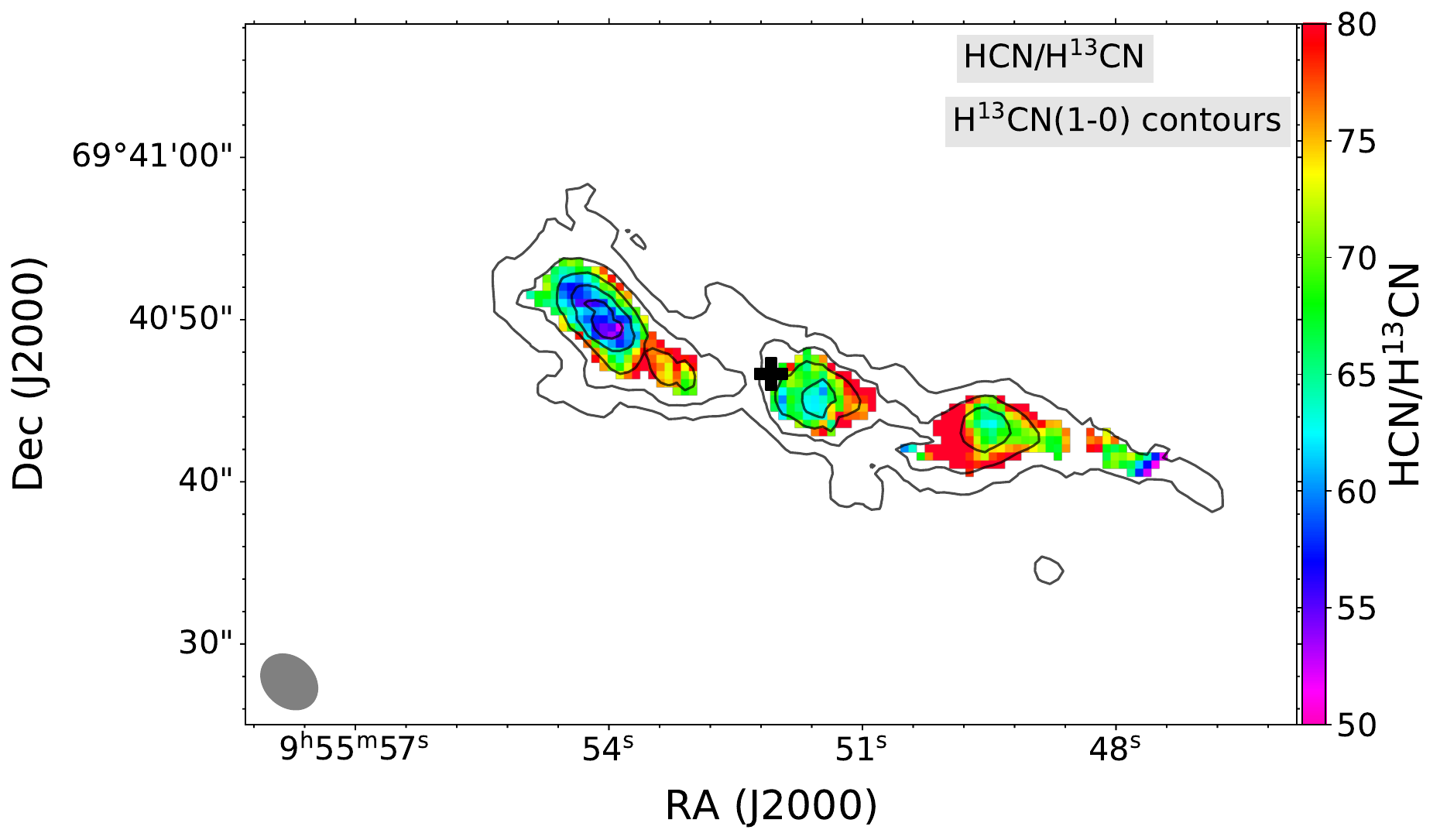}
\includegraphics[scale=0.27]{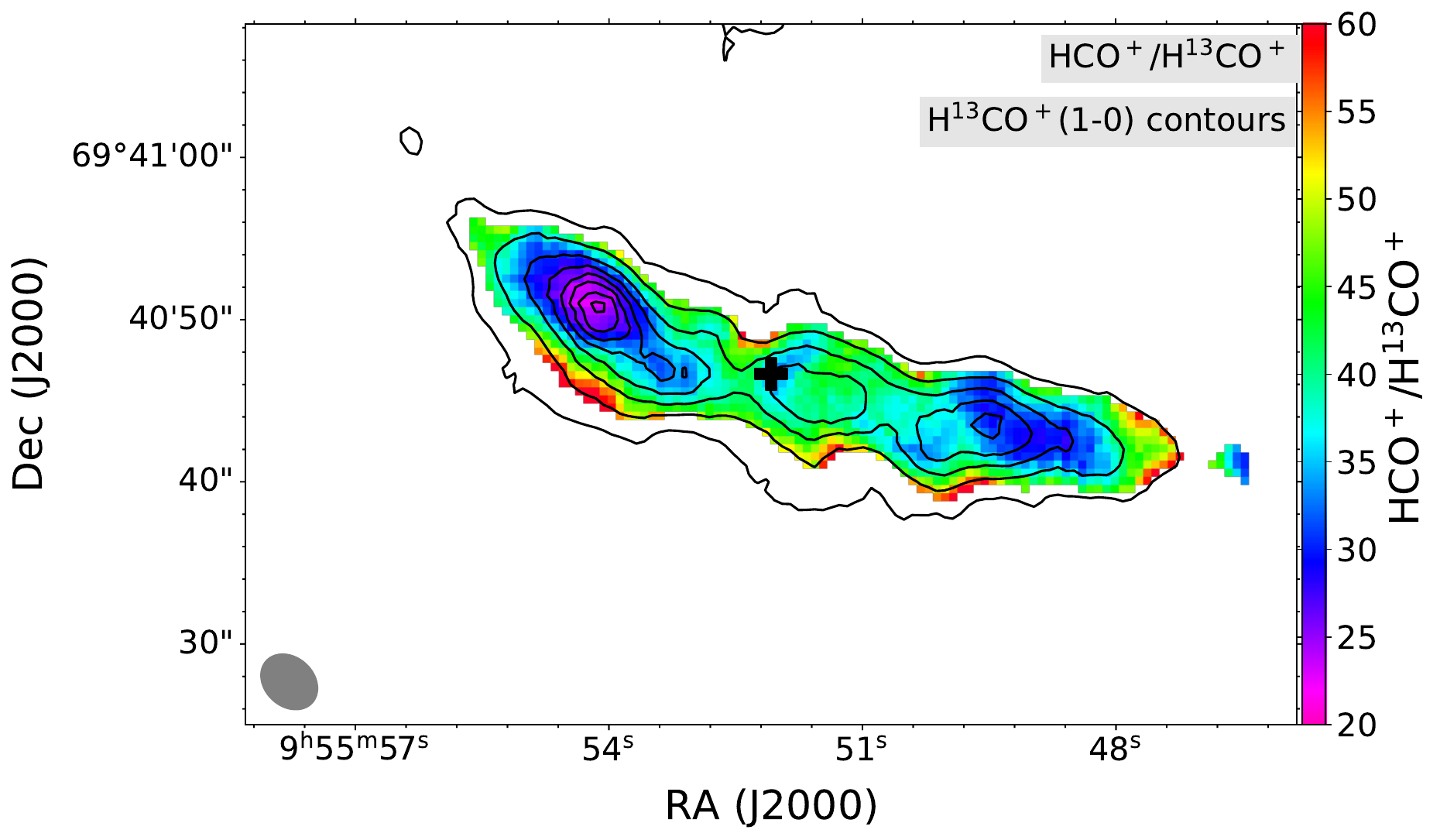}
\includegraphics[scale=0.27]{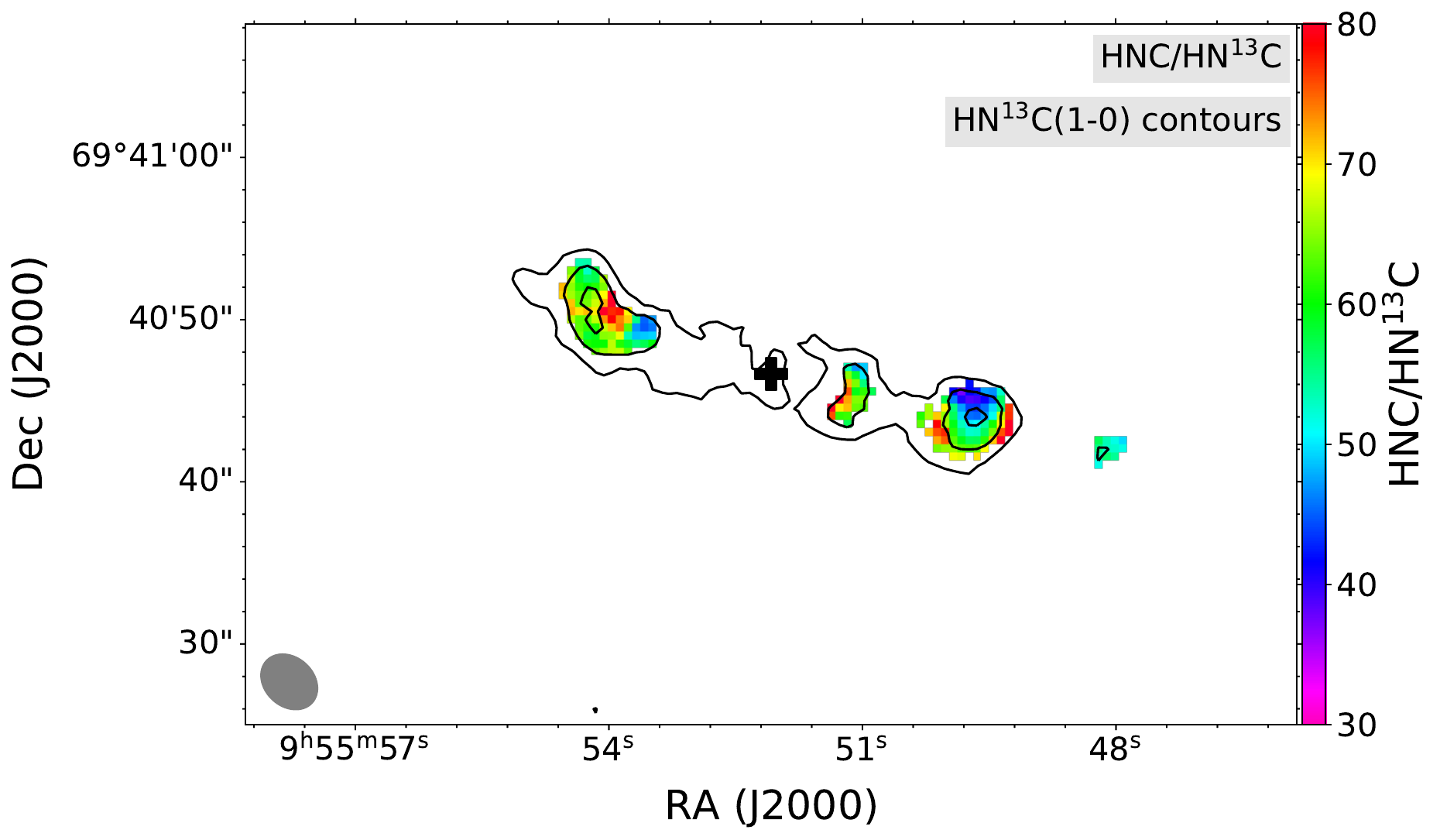}
\caption{Line ratio maps of HCN/H$^{13}$CN, HCO$^+$/H$^{13}$CO$^+$, and HNC/HN$^{13}$C. The contours  levels are same as Figure \ref{fig:isotopicline}. }
\label{fig:isotopic ratios}
\end{figure*}

\begin{figure*}
\includegraphics[scale=0.26]{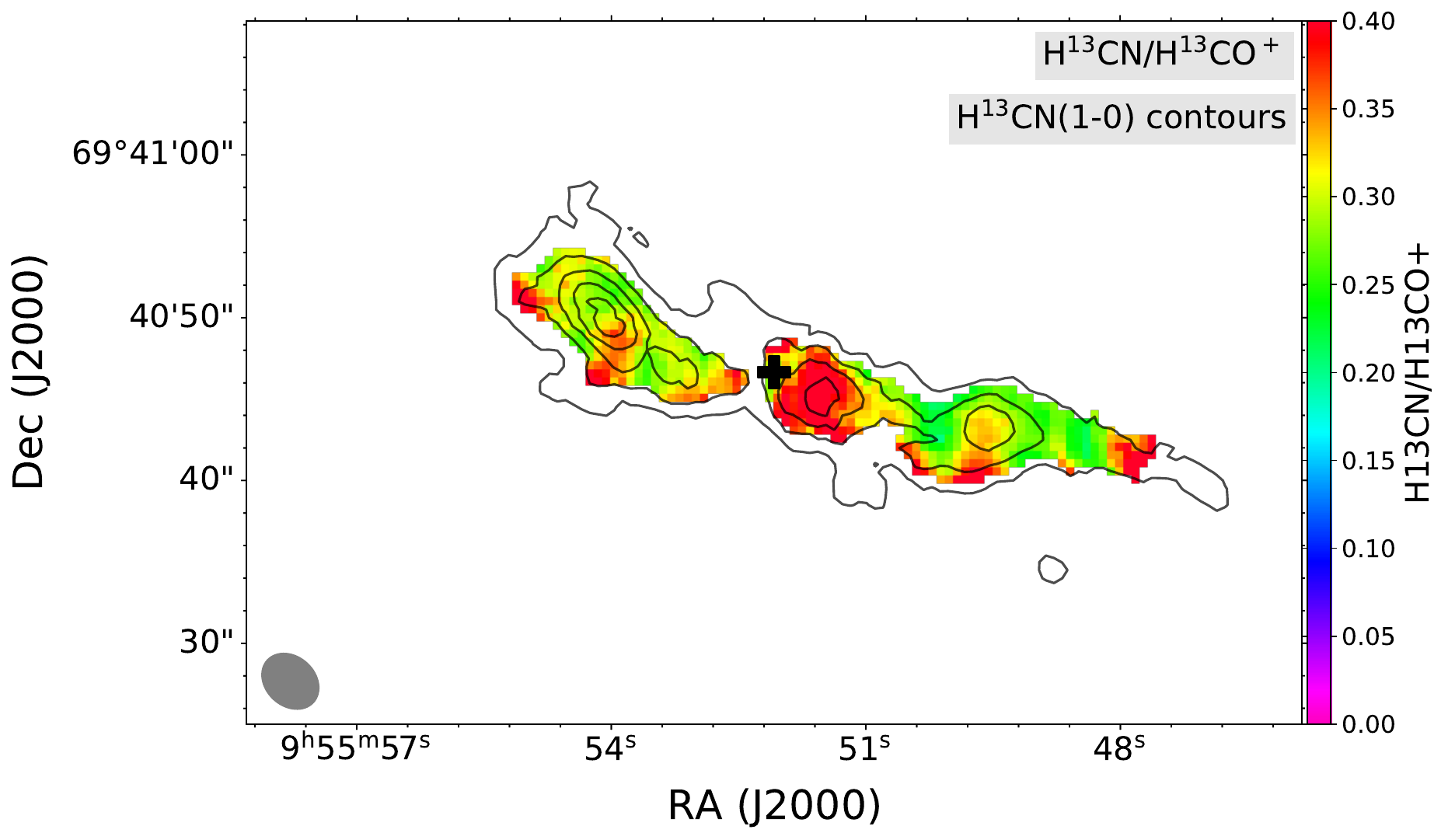}
\includegraphics[scale=0.26]{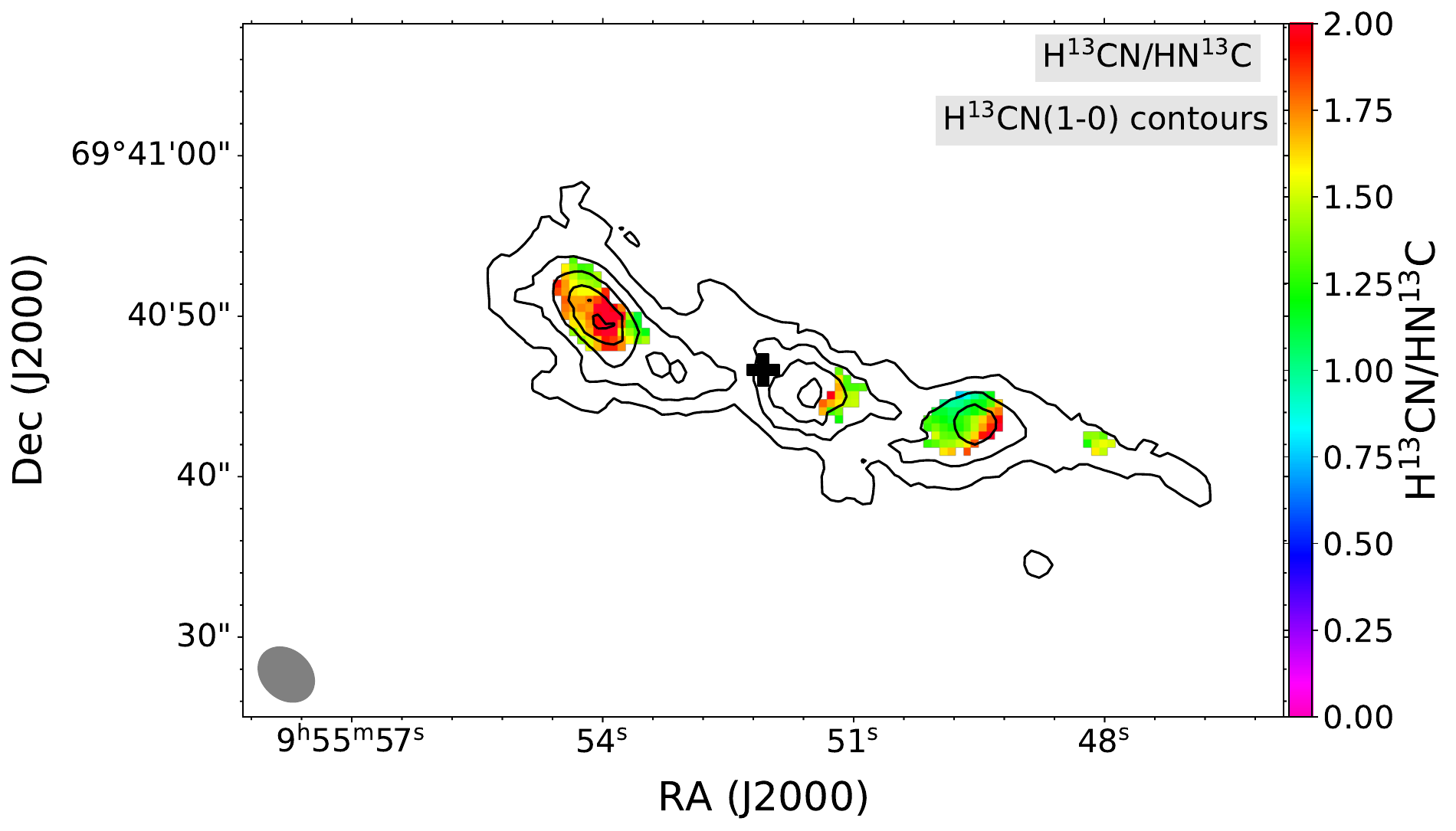}
\caption{Line ratio maps of H$^{13}$CN/H$^{13}$CO$^+$ and H$^{13}$CN/HN$^{13}$C. Two panels are overlaid with H$^{13}$CN contours. The contours levels are same as Figure  \ref{fig:isotopicline}.}
\label{fig:R_H13CN_H13COP_HN13C}
\end{figure*}


\section{The line ratio}

\subsection{Ratios between HCN, HCO$^+$, HNC and their isotopologues} \label{Appendix:line ratio}

In the following, we measure isotopologue ratio maps of HCN/H$^{13}$CN, HCO$^+$/H$^{13}$CO$^{+}$, and HNC/HN$^{13}$C. Although dense gas tracers are detected over large areas, only the inner, bright regions could yield useful constraints on the isotopologue line ratios. Therefore, we mainly concentrate on the brightest region to analyze these ratios (see Figure \ref{fig:isotopic ratios}). 
The ratios are calculated by taking the ratio of their velocity-integrated intensity and masking the pixels with 6$\sigma$ signal-to-noise level. This part focuses on the triple-peak regions (E1, W1 and W2). The uncertainty of the velocity-integrated intensity, resulting in a typical uncertainty of  intensity ratio is about 14\%. 

HCO$^+$/H$^{13}$CO$^{+}$ ratios are significantly lower than HCN/H$^{13}$CN and HNC/HN$^{13}$C, considering the relative range of ratios observed. It is consistent with the results from the single-dish data \citep{Li2022}.
HCN/H$^{13}$CN line ratio map shows the lowest value towards E1 position with 55$\pm$5, while it is 70$\pm$5 at W1 and W2 positions, although the values are high at the edge of the peak emissions. For HCO$^+$/H$^{13}$CO$^{+}$, gradients are seen from the center to the NE lobe and the SW lobe, with a low ratio ($<$30). It also shows the lowest ratio towards E1 position. While the lowest ratio ($<$40) of HNC/HN$^{13}$C associates with W2 position. 
Furthermore, HNC is sensitive to the high gas temperature, which could reduce the HCN abundance towards W2 positions in the outer regions \citep{Weiss2001}.
These three isotopologues ratios are higher than the observed CO/$^{13}$CO \citep{Kikumoto1998} and those of the other starburst galaxies of NGC 253 \citep{Meier2015} and NGC 4945 \citep{Henkel2018}. The reason maybe due to the isotopic abundance variations.






\subsection{H$^{13}$CN/H$^{13}$CO$^{+}$ and H$^{13}$CN/HN$^{13}$C}

Here we focus on the H$^{13}$CN/H$^{13}$CO$^{+}$, and H$^{13}$CN/HN$^{13}$C line ratios to further constrain the dense gas properties. H$^{13}$CN, H$^{13}$CO$^{+}$ and HN$^{13}$C have the advantage of being optically thin avoiding the opacity effects, which provide a better choice than HCN and HCO$^+$ and HNC (1-0) to derive HCN/HCO$^+$ and HCN/HNC abundance ratio.

The effective critical density of H$^{13}$CO$^{+}$ (1-0) is nearly a factor of five lower than H$^{13}$CN (1-0) \citep{Jimenez2017}. Therefore, the low H$^{13}$CN/H$^{13}$CO$^{+}$ ratio in the range of 0.2-0.4 are expected in moderate gas density (n$_{\rm H_2}$ $\sim$ 10$^5$ cm$^{-3}$). The highest value towards W1 might indicate that the  
gas density is relatively low there. On the other hand, H$^{13}$CO$^{+}$ is sensitive to the ionization fraction (electron abundance) \citep{Maret2007}, considering the potentially enhanced UV radiation and particularly the cosmic-rays in stardust regions, it will increase the H$^{13}$CN/H$^{13}$CO$^{+}$ ratios toward W1 position, as shown in Figure \ref{fig:R_H13CN_H13COP_HN13C}. H$^{13}$CN/H$^{13}$CO$^{+}$ line ratio in M 82 is lower by a factor of $\sim$ 2 than the measurements in NGC 253 \citep{Meier2015} and NGC 4945 \citep{Henkel2018}, which might be due to M 82 being a more evolved starburst \citep{Aladro2011b}.


HCN/HNC line ratio can reflect starburst evolution with gas phase chemistry in molecular clouds \citep{Aalto2002}. HNC/HCN ratio is subject to temperature variations \citep{Schilke1992,Meier2005} and it is expected to increase when the gas temperature becomes warmer, because HNC is destroyed in the hot gas. H$^{13}$CN/HN$^{13}$C ratio could be used to derive HCN/HNC abundance ratio.
Although H$^{13}$CN/HN$^{13}$C line ratio does not show a significant variation trend in M 82, the lowest value is observed towards W2 position, which is consistent with the relatively small HNC/HN$^{13}$C line ratio. \citep{Aalto2002} suggest that the HNC abundance becomes less dependent on temperature in PDR chemistry. Therefore, the gas temperature might affect HNC (1-0) emission, while it might not be the dominant mechanism.


\bibliography{sample631}{}
\bibliographystyle{aasjournal}



\end{document}